\documentclass[12pt]{iopart}
\usepackage{graphicx}

\newcommand{\be}{\begin{eqnarray}}
\newcommand{\ee}{\end{eqnarray}}
\newcommand{\ba}{\begin{array}}
\newcommand{\ea}{\end{array}}
\newcommand{\no}{\nonumber}

\begin{document}
\title[Replica analysis of partition-function zeros in SG models]
{Replica analysis of partition-function zeros in spin-glass models}
\author{Kazutaka Takahashi}
\address{Department of Physics, 
Tokyo Institute of Technology, Tokyo 152-8551, Japan}
\begin{abstract}
 We study the partition-function zeros in mean-field spin-glass models.
 We show that the replica method is useful to find  
 the locations of zeros in a complex parameter plane.
 For the random energy model, 
 we obtain the phase diagram in the plane 
 and find that there are two types of distribution of zeros:  
 two-dimensional distribution within a phase  
 and one-dimensional one on a phase boundary. 
 Phases with a two-dimensional distribution are 
 characterized by a novel order parameter defined 
 in the present replica analysis.
 We also discuss possible patterns of distributions
 by studying several systems.
\end{abstract}

\maketitle

\section{Introduction}

 Spin-glass (SG) systems are 
 known to show exotic thermodynamic properties
 and have been studied for many years~\cite{MPV, Nishimori, MM}.
 The existence of many quasi-stationary states 
 leads to a nontrivial critical behavior at low temperatures.
 Other systems such as structural glasses and 
 other theories like information processing
 share these properties.
 Therefore, a more general understanding
 is required for this randomness-induced transition.

 As a general method to understand the critical phenomena 
 of statistical mechanical systems, 
 Lee and Yang proposed to study the zeros of 
 the partition function~\cite{YL, LY}.
 Since the partition function is positive by definition, 
 the zeros are located in the unphysical region 
 of the complex parameter plane.
 In finite systems, 
 zeros are far from the real axis.
 However, as we make the system size large,  
 they reach a point on the real axis at the thermodynamic limit.
 If the partition function goes to zero
 at a certain point in parameter space, 
 the corresponding thermodynamic functions are singular, and  
 this point is identified as the phase transition point.
 For example, in the pure ferromagnetic Ising model, 
 the circle theorem states that 
 the zeros form a unit circle on 
 the complex fugacity plane~\cite{LY, NO}.
 In the same way, 
 we can consider zeros of the partition function 
 not only for complex fields  
 but also for other complex parameters.
 For example, the zeros in the complex temperature plane are 
 known as the Fisher zeros~\cite{Fisher}.

 While there are many studies on 
 the partition-function zeros for pure systems, 
 relatively few results exist for random ones.
 The zeros of the $\pm J$ Ising model were examined numerically 
 in \cite{ON, BL, DL, MNH} for finite-dimensional systems
 and in \cite{MMNOS} on the Bethe lattice. 
 The results show that 
 the zeros are distributed in the complex plane 
 in a very different manner compared to pure systems.
 That is, they are distributed two dimensionally 
 rather than one dimensionally.
 The two-dimensional distribution approaches the real axis 
 as the system size is made large,
 which is considered to be the onset of the SG phase transition.
 Therefore, it is important to understand how such a distribution is 
 formed in the SG systems.
 However, most of the previous studies were done numerically, and 
 we need a reliable analytical method.

 As a first step to understand the distribution of zeros 
 in SG systems analytically, 
 we propose a new method to use replicas 
 for mean-field SG models.
 The theory of spin glasses has been well understood 
 in terms of mean-field models with infinite-range interaction.
 It was found that  
 the replica method is useful for analytic evaluation of 
 the average free energy, 
 and replica symmetry breaking (RSB) corresponds to 
 the SG state~\cite{Parisi1, Parisi2, Parisi3}.
 The advantage of studying mean-field models 
 is that analytical calculations are possible, 
 and we can closely study how such a SG state is obtained.
 Therefore, it would be useful to study the zeros of 
 the partition function in mean-field SG models. 

 As a matter of fact, the analytical result is already available 
 for the simplest SG model, 
 the random energy model (REM)~\cite{Derrida1, Derrida2, GM}.
 In this model, the distribution of zeros 
 in the complex temperature plane 
 was calculated analytically in \cite{Derrida3}
 and confirmed numerically in \cite{MP1}.
 The case of complex magnetic field was also studied 
 both analytically~\cite{MP2} and numerically~\cite{MP3}.
 The method used in~\cite{Derrida3, MP2} was to calculate 
 the density of states for a given energy and 
 is very specific to the REM.
 In fact, the spin degrees of freedom have not been treated,  
 and it is hard to extend the calculation to other spin systems.
 Therefore, we propose a more general and systematic method 
 using replicas 
 so that the extension to other models is possible in principle.
 The use of the replica method has great advantages 
 since many technical methods have been established, 
 and we can utilize various properties obtained in the previous studies.

 The organization of this paper is as follows.
 In section~\ref{zeros}, we give a brief review of 
 partition-function zeros.
 Throughout this paper, we treat the REM and its variants.
 The model and the method of calculation are described 
 in section~\ref{model}.
 Then, in section~\ref{remb}, 
 we consider the Fisher zeros of the REM 
 and re-derive the result obtained in \cite{Derrida3}.
 Most of our main ideas and the essence of the calculations 
 are contained in this section.
 The following sections are the applications of the method 
 to several models.
 We treat the generalized REM (GREM)~\cite{Derrida4, DG1, DG2, OTT}
 in section~\ref{gremb}
 to study the effect of the RSB for the distribution of zeros.
 Also, to study the first-order transition, 
 we consider the REM with ferromagnetic interaction 
 in section~\ref{ferro}.
 The system in magnetic fields is considered to 
 find the Lee-Yang zeros in \ref{mag}. 
 Finally, in the last section~\ref{conc}, 
 we give conclusions and discuss issues for further study.

\section{Partition-function zeros}
\label{zeros}

 We give a brief introduction of the partition-function zeros 
 to derive several formulae used in the following sections. 
 We treat Ising spin systems in a magnetic field $h$ 
 and the Hamiltonian is generally written as 
\be
 H=H_0-h\sum_{i=1}^N S_i,
\ee
 where $S_i$ is the spin variable on site $i$, 
 $N$ the number of spins and
 $H_0$ is the field-independent part of the Hamiltonian.
 The partition function can be expressed by using 
 $y= e^{2\beta h}$ as 
\be
 Z = \Tr e^{-\beta H_0+\beta h\sum_{i=1}^NS_i}
 = e^{-N\beta h}
 \left(a_0+a_1y+a_2y^2+\cdots +a_Ny^N\right),
\ee
 where $\beta$ is the inverse temperature and 
 $a_0, a_1, \cdots, a_N$ are coefficients 
 determined by $H_0$.
 Apart from the overall factor $e^{-N\beta h}$,  
 this function is a polynomial of the $N$th degree in $y$ 
 and is factorized as 
\be
 Z = a_N e^{-N\beta h}\prod_{i=1}^N(y-y(i)), \label{fact}
\ee
 where $y(i)$ are the complex numbers and represent the zeros of 
 the partition function.
 Then, the free energy per spin
 $f=-(1/N\beta)\ln Z$ can be written as 
\be
 -\beta f = \frac{1}{N}\ln a_N -\beta h +
 \int dz_1 dz_2\, \rho_y(z_1,z_2) \ln(y-z),
 \label{f}
\ee
 where $y_1$ and $y_2$ denote the real and imaginary parts 
 of $y$, respectively, and the same is for $z_1$ and $z_2$. 
 The density of zeros $\rho_y(y_1,y_2)$ is introduced as 
\be
 \rho_y(y_1,y_2) = \frac{1}{N}\sum_{i=1}^N \delta(y-y(i)).
\ee
 Thus, the free energy can be written as an integral over 
 zeros in the complex $y$ plane.

 The expression of the free energy in terms of the density of zeros 
 indicates the following features 
 on the analyticity of thermodynamic functions.
 If the free energy (\ref{f}) has a singularity, 
 it should come from the integral at the point $z=y$.
 Since $y=e^{2\beta h}$ is real and 
 zeros are not on the real axis for finite systems, 
 the phase transition indicated by the singularity 
 is realized only at the thermodynamic limit.
 We can observe how zeros approach the real axis by increasing 
 the system size $N$.

 Our analysis in the following sections is based on the formula
\be
 \rho_y(y_1,y_2) = \frac{1}{2\pi N}\left(
 \frac{\partial^2}{\partial y_1^2}+\frac{\partial^2}{\partial y_2^2}
 \right)\ln|Z e^{N\beta h}|, \label{rho}
\ee
 which is derived from the expression (\ref{fact})~\cite{Derrida3}.
 Here, the magnetic field in the partition function
 takes a complex value.
 That is, the density of zeros can be obtained from 
 the absolute value of the partition function 
 with complex-valued fields.

 We mainly consider in the following the Fisher zeros,
 the partition-function zeros in the complex temperature plane.
 Assuming that the partition function is factorized with respect to 
 $\beta$ as 
\be
 Z(\beta) = C\prod_{i}(\beta-\beta(i)),
\ee
 we can write
\be
 & & \ln Z(\beta) = \ln C + N\int dz_1dz_2\,
 \rho_\beta(z_1,z_2)\ln(\beta-z), \label{lnZ} \\
 & & \rho_{\beta}(\beta_1,\beta_2) 
 = \frac{1}{2\pi N}
 \left(
 \frac{\partial^2}{\partial\beta_1^2}
 +\frac{\partial^2}{\partial\beta_2^2}
 \right)\ln |Z(\beta=\beta_1+i\beta_2)|,
 \label{rhob}
\ee
 where $\rho_{\beta}$ is the density of zeros in the complex-$\beta$ plane.
 We note that $\rho_\beta$ is not normalized to unity.  
 This is because the partition function is 
 a polynomial of infinite degree in $\beta$ even for finite $N$. 

\section{Mean-field SG models and replica method}
\label{model}

\subsection{Model}

 We treat the REM
 which is represented by randomly distributed energy levels.
 For a given set of energies $(E_1,E_2\cdots, E_{2^N})$,
 the partition function is written as 
\be
 Z = \sum_{i=1}^{2^N} e^{-\beta E_i}. \label{Z}
\ee
 Each energy level is taken from the Gaussian distribution 
\be
 {\rm P}_{\rm E}(E_i) = \frac{1}{\sqrt{\pi NJ^2}}\exp\left(
 -\frac{E_i^2}{NJ^2}
 \right). \label{gauss}
\ee
 It is well known that this model is equivalent to 
 the $p\to\infty$ limit of the $p$-body interacting Ising spin model 
 whose Hamiltonian is given by 
\be
 & & H = -\sum_{i_1<i_2<\cdots<i_p}^{N}J_{i_1i_2\cdots i_p}
 S_{i_1}S_{i_2}\cdots S_{i_p},  \label{pspin}
\ee
 where the probability distribution of 
 the interaction $J_{i_1\cdots i_p}$ is Gaussian as
\be
 & & {\rm P}_{\rm J}(J_{i_1\cdots i_p})=
 \sqrt{\frac{N^{p-1}}{\pi p! J^2}}
 \exp\left\{
 -\frac{N^{p-1}}{p!J^2}
 \left(J_{i_1i_2\cdots i_p}\right)^2
 \right\}.
\ee

 The REM is solved exactly, and we can find a phase transition 
 between the paramagnetic (P) and SG phases 
 at $\beta=\beta_{\rm c}$ where
\be
 \beta_{\rm c}J=2\sqrt{\ln 2}. \label{betac}
\ee
 At temperatures lower than $T_{\rm c}=1/\beta_{\rm c}$,
 the system freezes to its ground state and 
 the entropy goes to zero.
 This SG state is obtained by the one-step RSB (1RSB) ansatz
 in the replica formalism.

\subsection{Replica method}

 In order to treat disordered systems, 
 it is useful to take an ensemble average over disorder parameters.
 The self-averaging property ensures that one can study 
 the thermodynamic limit of the system 
 by using the averaged free energy.
 Then, the replica method is useful to calculate the average.
 As we described in the previous section, 
 in order to obtain the density of zeros, 
 we need to calculate $[\ln |Z|]$ 
 where the square bracket denotes the average.
 The main idea of this paper is to use the formula
\be
 \ln |Z| = \lim_{n\to 0}\frac{|Z|^{2n}-1}{2n}.
\ee
 Compared to the standard procedure to treat 
 $[Z^n]$ for the free energy $-\beta F = \ln Z$, 
 we see that the number of replicas is doubled: 
 $[|Z|^{2n}]=[Z^nZ^{*n}]$.
 Then, the resultant SG order parameter $Q_{ab}$
 is represented by a $2n\times 2n$ matrix.
 With complex parameters, the elements of 
 the order-parameter matrix can take a complex value.
 However, in the case of the REM, they are real, 
 which allows us to obtain the result easily. 

 Since the replica space is doubled and the parameter is complex, 
 we can find new phases which do not exist in systems with real parameter.
 In such a phase, it is shown in the following section 
 that the factorization property 
\be
 [|Z|^{2n}] = [Z^n][Z^{*n}] \label{zz}
\ee
 does not hold.
 In fact, this property is closely related to 
 whether the two-dimensional distribution of zeros 
 appears or not.

\section{Fisher zeros of the REM}
\label{remb}

 As a first example, we consider the REM with complex-$\beta$.
 In this case, the density of zeros has been obtained 
 analytically by Derrida in 1991~\cite{Derrida3}.
 His method corresponds to the microcanonical calculation 
 in some sense  
 since the density of states as a function of energy is  
 calculated to obtain the canonical partition function.
 Here, we re-derive the result by the canonical replica calculation.

\subsection{Replica method}

 For a given $(E_1,E_2,\cdots, E_{2^N})$, 
 the partition function is defined as (\ref{Z}).
 Then, we can write  
\be
 |Z|^{2n} = \sum_{\{i_a\}}\sum_{\{i'_a\}}
 \exp\left\{
 -\beta\sum_{j=1}^{2^N}n_j(\{i_a\})E_j
 -\beta^*\sum_{j=1}^{2^N}n_j(\{i'_a\})E_j
 \right\},
\ee
 where 
\be
 n_j(\{i_a\})=\sum_{a=1}^n\delta_{i_aj},
\ee
 and 
 the configuration sums are taken over 
 $\{i_a\}=(i_1,i_2,\cdots, i_n)$ and
 $\{i'_a\}=(i'_1,i'_2,\cdots, i'_n)$ 
 with each $i_a$ and $i'_a$ taking $1,2,\cdots ,2^N$.
 Taking the average (\ref{gauss}), we obtain 
\be
 [|Z|^{2n}]
 = \sum_{\{i_a\}}\sum_{\{i'_a\}}
 \exp\left\{
 \frac{N\beta^2J^2}{4}\sum_{ab}q_{ab}
 +\frac{N\beta^{*2}J^2}{4}\sum_{ab}q'_{ab}
 +\frac{N|\beta|^2J^2}{2}\sum_{ab}\tilde{q}_{ab}
 \right\}. \no\\
\ee
 Here, three order parameters are introduced as 
\be
 & & q_{ab}=\delta_{i_ai_b}, \\
 & & q'_{ab}=\delta_{i'_ai'_b}, \\
 & & \tilde{q}_{ab}=\delta_{i_ai'_b},
\ee
 which can be written in a $2n\times 2n$ matrix form as 
\be
 Q = \left(\ba{cc} q & \tilde{q} \\ \tilde{q}^T & q' \ea\right).
\ee
 The $n\times n$ matrix $q$ represents the overlap between 
 configurations in $\{i_a\}$, 
 $q'$ in $\{i'_a\}$ and 
 $\tilde{q}$ in $\{i_a\}$ and $\{i'_a\}$.
 Each matrix element takes 0 or 1.
 At the thermodynamic limit $N\to\infty$, 
 the saddle-point configurations dominate 
 the sum, and we can write 
\be
 [|Z|^{2n}]=\exp\left\{N
 \left(s_q +\frac{\beta^2J^2}{4}\sum_{ab}q_{ab}
 +\frac{\beta^{*2}J^2}{4}\sum_{ab}q'_{ab}
 +\frac{|\beta|^2J^2}{2}\sum_{ab}\tilde{q}_{ab}
 \right)\right\}, \no\\
\ee
 where $Ns_q$ is the entropy function 
 for configurations to give the saddle-point solution.

 Since $[|Z|^{2n}]$ is real, the saddle-point solution 
 must satisfy the relation
\be
 \sum_{ab}q_{ab}=\sum_{ab}q'_{ab}. \label{qq'}
\ee
 Then, 
\be
 [|Z|^{2n}]=\exp\left\{N\left(
 s_q +\frac{J^2}{2}(\beta_1^2-\beta_2^2)\sum_{ab}q_{ab}
 +\frac{J^2}{2}(\beta_1^2+\beta_2^2)\sum_{ab}\tilde{q}_{ab}
 \right)\right\}. \label{Z2n}
\ee
 As possible saddle-point solutions, 
 we can consider the replica symmetric (RS) and 1RSB ones.
 Among them, we choose the solution that gives the maximum $\ln|Z|$.
 We also impose the condition that 
 the thermodynamic entropy must be nonnegative. 
 The definition of the entropy for complex-$\beta$ is nontrivial, 
 and we discuss it in the next subsection. 

\subsection{Entropy for complex temperature}
\label{entropy}

 In the standard theory of statistical mechanics, 
 the partition function is written as 
\be
 Z = \int dE\, e^{-\beta E+S(E)}.
\ee
 The entropy $S(E)$ is defined as the logarithm of the number of states 
 for a given energy $E$.
 At the thermodynamic limit, the saddle point
 $\beta=dS/dE$ contributes to the energy integral.
 When $\beta$ is complex as $\beta=\beta_1+i\beta_2$, 
 the saddle-point energy is also given by a complex number.
 We change the integral path so that the path 
 crosses the saddle point.
 For an obtained saddle point $E=E_1+iE_2$, 
 we can write 
\be
 |Z|^2 = e^{-2\beta_1 E_1+2\beta_2 E_2+2S(E_1,E_2)},
\ee
 where we introduce the real function $S(E_1,E_2)$.
 We write for $\Phi=\ln|Z|$, 
\be
 \Phi(\beta_1,\beta_2)=-\beta_1 E_1+\beta_2 E_2+S(E_1,E_2).
 \label{phis}
\ee
 Thus, $\Phi(\beta_1,\beta_2)$ is obtained from $S(E_1,E_2)$ by 
 the Legendre transformation, and vice versa. 
 The parameters $\beta_{1,2}$ and $E_{1,2}$ are related as 
\be
 E_1=-\frac{\partial \Phi}{\partial \beta_1}, \qquad
 E_2=\frac{\partial \Phi}{\partial \beta_2}.
\ee

 $S(E_1,E_2)$ defined here 
 is considered a natural generalization of the ordinary 
 entropy for real parameters.
 It is real and we employ the nonnegative 
 condition of $S(E_1,E_2)$ when we determine 
 the effective domain of a phase.

\subsection{Saddle-point solutions and phase diagram}

\begin{center}
\begin{figure}[htb]
\begin{minipage}[h]{0.5\textwidth}
\begin{center}
\includegraphics[width=0.8\columnwidth]{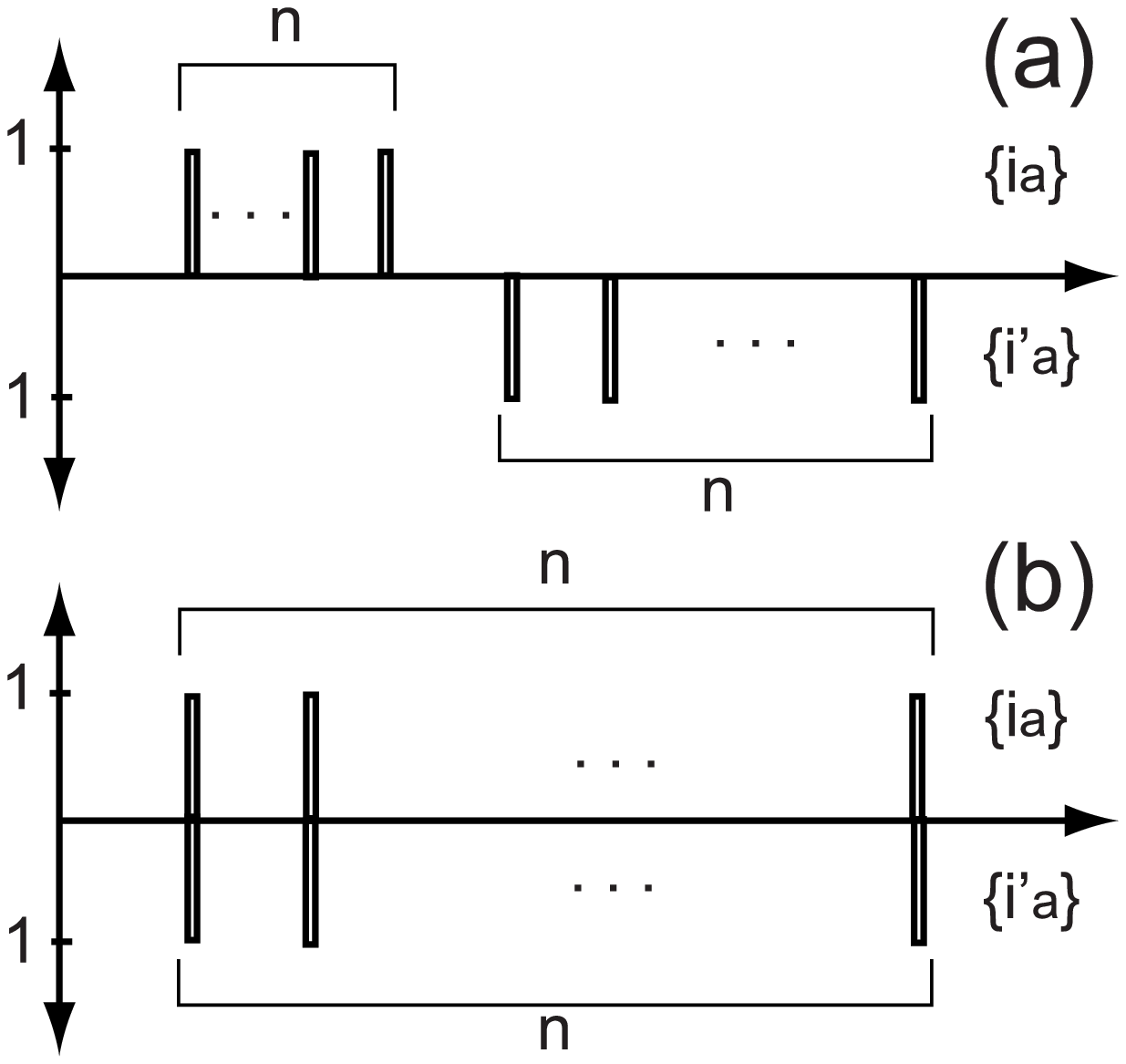}
\caption{
 RS saddle-point configurations (\ref{RS}). 
 The horizontal axis represents the index of configurations running 
 from 1 to $2^N$.
 The vertical axis represents the number of each configurations. 
 (a) P1 phase (\ref{p1}) and (b) P2 (\ref{p2}). 
 }
\label{rs}
\end{center}
\end{minipage}
\begin{minipage}[h]{0.5\textwidth}
\begin{center}
\includegraphics[width=0.8\columnwidth]{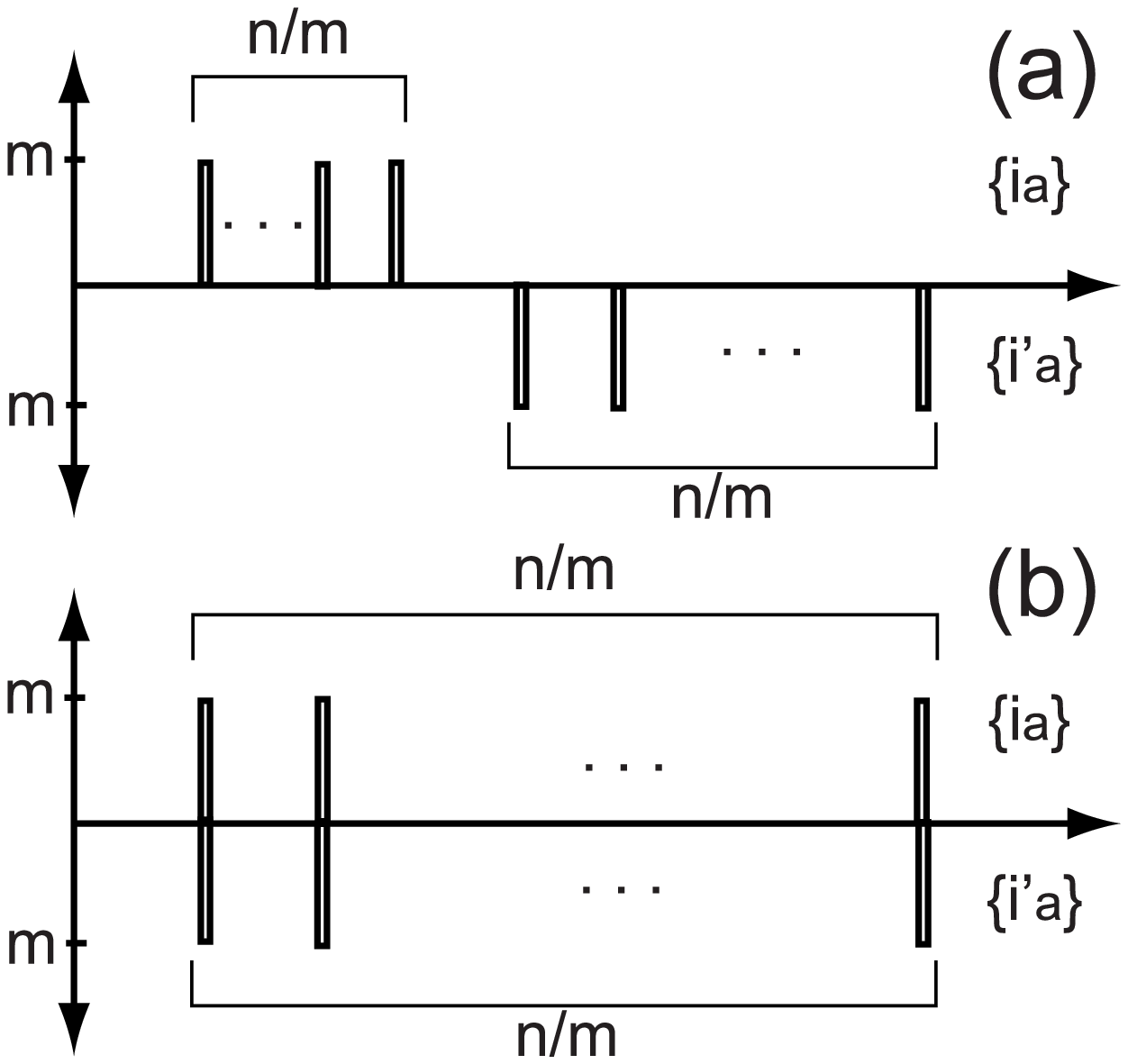}
\caption{1RSB saddle-point configurations (\ref{1rsb-q}). 
 (a) SG phase with $\tilde{q}=0$ and (b) that with $\tilde{q}\ne 0$. 
}
\label{rsb}
\end{center}
\end{minipage}
\end{figure}
\end{center}

 Now we consider the saddle-point solutions for (\ref{Z2n}).
 Since $q_{aa}=q'_{aa}=1$ by definition, 
 we see that the simplest solution is given by
\be
 q_{ab}=q'_{ab}=\delta_{ab}. \label{RS}
\ee
 This result is known as the ordinary RS solution and 
 corresponds to the P phase.
 However, (\ref{RS}) does not completely specify the saddle point, 
 and we have an additional order-parameter matrix $\tilde{q}$ 
 which represents overlaps between the configurations 
 $\{i_a\}$ and $\{i'_a\}$.
 If there is no correlation between them, we have
\be
 \tilde{q}_{ab}=0,\label{p1}
\ee
 which we call the P1 phase.
 It is also possible to consider 
 \footnote{More generally, 
 this should be written as 
 $\sum_{ab}\tilde{q}_{ab}=n$
 since (\ref{Z2n}) depends only on the sum.}
\be
 \tilde{q}_{ab}=\delta_{ab}, \label{p2}
\ee
 which means that the configurations are completely the same:
 $\{i_a\}=\{i'_a\}$.
 We call the corresponding phase the P2 one.
 These solutions can be graphically expressed in figure~\ref{rs}.
 $n$ states are chosen from $2^N$-configurations.
 Figure~\ref{rs}(a) represents the P1 phase and 
 \ref{rs}(b) the P2 one.
 In principle, we can consider intermediate states 
 between P1 and P2 phases
 such that $0<\sum_{ab}\tilde{q}_{ab}<n$, 
 but they do not give the maximum $\ln |Z|$.

 In the P1 and P2 phases, 
 the numbers of states giving the saddle-point value of $Q$
 are respectively given by 
\be
 e^{Ns_q} \sim \left\{\ba{cc}
 e^{2Nn\ln 2} & {\rm P1} \\
 e^{Nn\ln 2} & {\rm P2}
 \ea\right..
\ee
 We have for $\phi=[\ln|Z|]/N$ 
\be
 \phi = 
 \left\{\ba{ll}
 \ln 2+\frac{1}{4}(\beta_1^2-\beta_2^2)J^2 & {\rm P1} \\
 \frac{1}{2}\ln 2+\frac{1}{2}\beta_1^2J^2 &
 {\rm P2}
 \ea\right..
\ee
 The entropy in each phase is calculated from (\ref{phis}).
 We obtain the entropy per spin 
\be
 s = \left\{\ba{ll}
 \ln 2-\frac{1}{4}(\beta_1^2-\beta_2^2)J^2 & {\rm P1} \\
 \frac{1}{2}\ln 2-\frac{1}{2}\beta_1^2J^2 & {\rm P2}
 \ea\right..
\ee
 The condition $s\ge 0$ gives us 
\be
 -\beta_1^2+\beta_2^2+\beta_{\rm c}^2 \ge 0 \label{nonn}
\ee
 in the P1 phase and 
\be
 |\beta_1| \le \beta_{\rm c}/2
\ee
 in the P2 phase.
 Here, the inverse critical temperature $\beta_{\rm c}$ is 
 defined as (\ref{betac}).

 Next, we treat the 1RSB SG solution which can be expressed 
 in figure~\ref{rsb}.
 We have for $q_{ab}$ and $q'_{ab}$
\be
 \sum_{ab}q_{ab}=\sum_{ab}q'_{ab}=mn,
 \label{1rsb-q}
\ee
 where $m$ is the Parisi breaking parameter.
 For the configuration in figure~\ref{rsb}(a), 
 there is no overlap between $\{i_a\}$ and $\{i'_a\}$ and
\be
 \tilde{q}_{ab}=0.
\ee
 For that in figure~\ref{rsb}(b), 
 $\tilde{q}_{ab}$ has the 1RSB structure and 
\be
 \sum_{ab}\tilde{q}_{ab}=mn.
 \label{1rsb-qt}
\ee
 Then, for each saddle point, 
\be
 \phi = \left\{\ba{ll}
 \frac{\ln 2}{m}
 +\frac{m}{4}(\beta_1^2-\beta_2^2)J^2 \\
 \frac{\ln 2}{2m}
 +\frac{m}{2}\beta_1^2J^2
 \ea\right..
\ee
 $m$ is optimized as 
\be
 m = \left\{\ba{l}
 \frac{\beta_{\rm c}}{\sqrt{\beta_1^2-\beta_2^2}} \\
 \frac{\beta_{\rm c}}{2|\beta_1|} 
 \ea\right.,
\ee
 and we obtain
\be
 \phi= \left\{\ba{l}
 \frac{J^2}{2}\beta_{\rm c}\sqrt{\beta_1^2-\beta_2^2} \\
 \frac{J^2}{2}\beta_{\rm c}|\beta_1|
 \ea\right..
\ee
 In both the cases, the entropy is shown to be zero.
 The former solution is always smaller than the latter one 
 and is discarded.

\begin{center}
\begin{figure}[htb]
\begin{minipage}[h]{0.5\textwidth}
\begin{center}
\includegraphics[width=1.0\columnwidth]{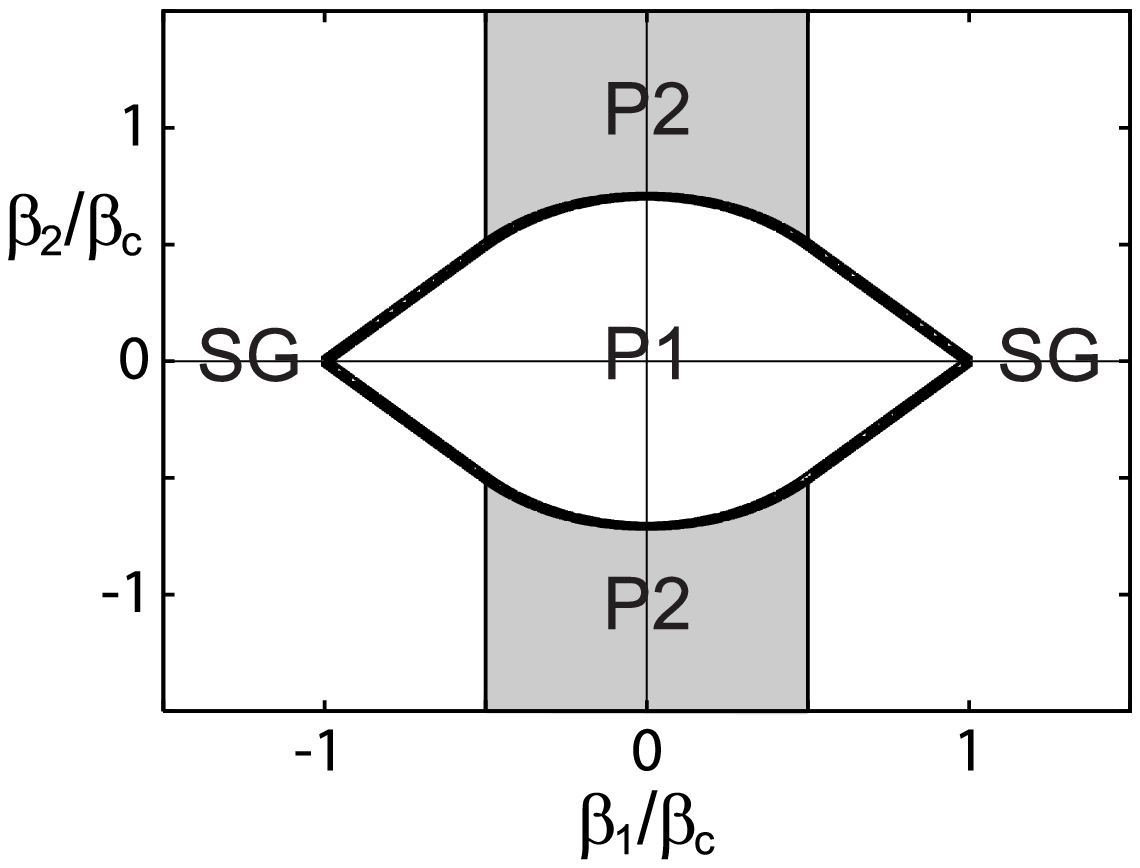}
\caption{Phase diagram and distribution of partition-function zeros 
 of the REM in the complex-$\beta$ plane.
 Zeros are distributed in the shaded area and on the bold lines.}
\label{rem}
\end{center}
\end{minipage}
\begin{minipage}[h]{0.5\textwidth}
\begin{center}
\includegraphics[width=1.0\columnwidth]{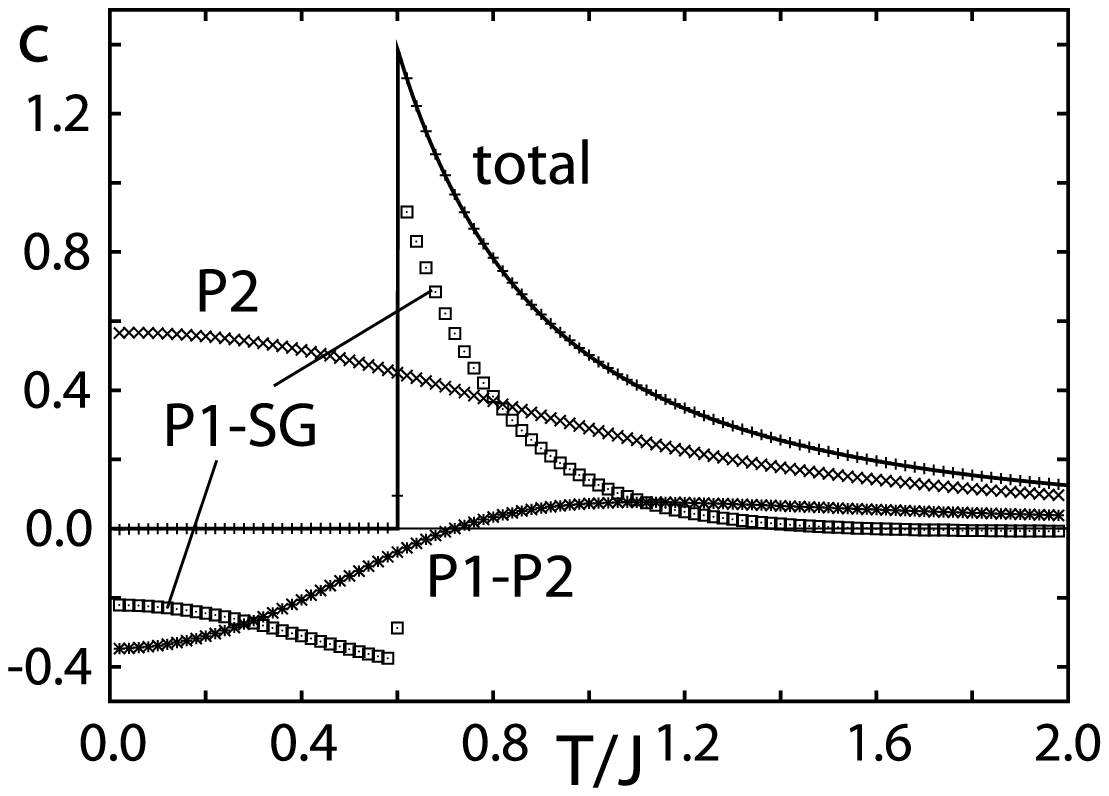}
\caption{The specific heat calculated from (\ref{cint}).
 The solid line represents the exact result 
 $c=(J^2/2T^2)\theta(T-T_{\rm c})$ where $T_{\rm c}=1/\beta_{\rm c}$.
}
\label{c}

\end{center}
\end{minipage}
\end{figure}
\end{center}

 We summarize the result as 
\be
 \phi = \left\{
 \begin{array}{lll}
 \frac{J^2}{4}(\beta_1^2-\beta_2^2+\beta_{\rm c}^2) & 
 (\beta_1^2-\beta_{\rm c}^2 \le \beta_2^2) & {\rm P1}  \\ 
 \frac{J^2}{2}\left(\beta_1^2+\frac{\beta_{\rm c}^2}{4}\right) &
 (|\beta_1| \le \beta_{\rm c}/2) & {\rm P2} \\
 \frac{J^2}{2}\beta_{\rm c}|\beta_1| & & {\rm SG} 
 \end{array}
 \right..
 \label{phi-rem}
\ee
 We compare these solutions to take the maximum value of $\phi$.
 The condition that P1 is larger than P2 is 
\be
 \beta_1^2+\beta_2^2 < \frac{\beta_{\rm c}^2}{2}, 
\ee
 and that P1 is larger than SG is 
\be
 (|\beta_1|-\beta_{\rm c})^2 > \beta_2^2.
\ee
 SG is always larger than P2.
 They are equal when $|\beta_1|=\beta_{\rm c}/2$.
 In conclusion, the phase diagram is obtained as in figure~\ref{rem}.

\subsection{Density of zeros}

 Now we calculate the density of zeros 
 by using formula (\ref{rhob}).
 The density of zeros in each phase is given by
\be
 \rho_\beta(\beta_1,\beta_2) = 
 \left\{\ba{ll}
 0 & {\rm P1} \\
 J^2/2\pi & {\rm P2} \\
 0 & {\rm SG} 
 \ea\right..
\ee
 In addition to this, zeros are distributed 
 on phase boundaries where the derivative of the free energy 
 shows a discontinuous change.
 On the P1-P2 boundary, that is,  
 $\beta_1^2+\beta_2^2 =\beta_{\rm c}/2$ and $0<|\beta_1|/\beta_{\rm c}<1/2$, 
 the density of zeros is calculated as 
\be
 \rho_{\beta}(\beta_1,\beta_2) 
 &=& \frac{1}{2\pi}\left(
 \frac{\partial^2}{\partial\beta_1^2}
 +\frac{\partial^2}{\partial\beta_2^2}\right)
 \left\{\phi_{\rm P1}\theta(\phi_{\rm P1}-\phi_{\rm P2})
 +\phi_{\rm P2}\theta(\phi_{\rm P2}-\phi_{\rm P1})
 \right\} \\ 
 &=& \frac{J^2}{2\pi}\frac{\beta_{\rm c}^2}{2}
 \delta(\beta_1^2+\beta_2^2-\beta_{\rm c}^2/2),
\ee
 where $\theta$ is the step function such that 
 $\theta(x)=1$ when $x$ is positive and 0 when negative.
 In the same way, on the P1-SG boundary
 $|\beta_1|+|\beta_2| =\beta_{\rm c}$ and 
 $1/2<|\beta_1|/\beta_{\rm c}<1$, 
\be
 \rho_{\beta}(\beta_1,\beta_2) 
 = \frac{J^2}{2\pi}\beta_2\delta(|\beta_1|-\beta_{\rm c}+|\beta_2|).
\ee
 On the P2-SG boundary 
 $|\beta_1|/\beta_{\rm c}=1/2$ and $1/2<|\beta_2|/\beta_{\rm c}$, 
 the boundary distribution of zeros is shown to be zero.
 We conclude that the density of zeros in the REM is given by 
\be
 \rho_{\rm \beta}(\beta_1,\beta_2) &=& \frac{J^2}{2\pi}
 \theta(\beta_{\rm c}/2-|\beta_1|)
 \theta(\beta_1^2+\beta_2^2-\beta_{\rm c}^2/2) 
 \no\\
 & & +\frac{J^2}{2\pi}\frac{\beta_{\rm c}^2}{2}
 \delta(\beta_1^2+\beta_2^2-\beta_{\rm c}^2/2)
 \theta(\beta_{\rm c}/2-|\beta_1|) \no\\
 & & 
 +\frac{J^2}{2\pi}\beta_2\delta(|\beta_1|+|\beta_2|-\beta_{\rm c})
 \theta(|\beta_1|-\beta_{\rm c}/2)
 \theta(\beta_{\rm c}-|\beta_1|).
\ee
 The result coincides with that in \cite{Derrida3}.

\subsection{Discussions}

 We have re-derived the density of zeros by using the replica method,
 which allows us to clarify 
 the following properties of the distribution:

\begin{itemize}

\item 
 Figures of the distribution of zeros can be viewed as 
 phase diagrams in the complex plane.
 There are two types of distributions for the density of zeros: 
 two-dimensional distribution within a phase and 
 one-dimensional one at a phase boundary.

\item 
 When $\tilde{q}_{ab}=0$,
 the two configurations $\{i_a\}$ and $\{i'_a\}$
 become independent of each other.
 Then, the average partition function 
 decouples as (\ref{zz}) and 
 $\phi$ is written as a sum 
 $\phi=\phi_0(\beta)+\phi_0(\beta^*)$.
 In this case, the density of state is shown to be zero.

\item 
 When $\tilde{q}_{ab}\ne 0$, 
 there are correlations between 
 two configurations $\{i_a\}$ and $\{i'_a\}$.
 Then, two-dimensional distributions of zeros 
 can be obtained.
 Since the correlation occurs
 only when the ensemble average is taken, 
 one can say that the two-dimensional distribution is specific
 to random systems.  
 In the case of the REM, 
 two configurations are completely identical: $\{i_a\}=\{i'_a\}$ 
 in the P2 and SG phases.

\item
 However, $\tilde{q}_{ab}\ne 0$ is not the sufficient condition 
 for two-dimensional distributions.
 In the REM, a two-dimensional distribution appears 
 in the P2 phase and not in the SG one. 
 Our result indicates that two-dimensional distributions are not 
 specific to the SG phase.
 In the SG phase of the REM, the system freezes to its ground state
 and the entropy goes to zero.
 In such a frozen phase, zeros do not appear.

\item 
 The two-dimensional distribution of zeros in the P2 phase 
 is located far from the real physical axis 
 and near the imaginary axis.
 On the imaginary axis, the Boltzmann factor is given by 
 $e^{-i\beta_2 H}$ and can take arbitrary values.
 Therefore, it is a natural result that 
 zeros appear around the imaginary axis.
 It is interesting to see that the zeros can appear 
 only when $\beta_2>\beta_{\rm c}/\sqrt{2}$ on the imaginary axis.
 Since $e^{-i\beta_2 H}$ is nothing but 
 the time evolution operator, 
 the result may be related to dynamical properties of the system.

\item 
 The one-dimensional distribution of zeros on the P1-SG boundary 
 meets with the real axis at the angle of $\pi/4$, and 
 the density of zeros is zero at 
 the phase transition point on the real axis. 
 These properties are characteristics of second-order phase 
 transitions~\cite{Abe}.
 We discuss these properties closely in section~\ref{ferro}.

\item To understand the role of each zero, 
 we calculate the specific heat from the obtained density of zeros.
 Using (\ref{lnZ}), we can write the specific heat as 
\be
 c = \int dz_1dz_2\,\rho_\beta(z_1,z_2)\frac{-\beta^2}{(\beta-z)^2}, 
 \label{cint}
\ee
 which allows us to write the function  
 as a sum of contributions from each phase and boundary.
 The result is plotted in figure~\ref{c}.
 The singularity at the transition point comes from 
 the zeros in the P1-P2 boundary.
 At large temperature, the main contribution comes from 
 the zeros in the P2 phase.
 In the SG phase lower than the transition point,
 the contributions from the boundaries give a negative specific heat.
 All contributions cancel out to give the zero specific heat.
 As we decrease the temperature, 
 zeros far from the real axis in the P2 phase
 become more important.

\item 
 In section~\ref{entropy}, 
 we have defined the entropy for complex temperature
 by using the Legendre transformation.
 It is not obvious whether
 we should impose the nonnegative condition to the entropy 
 since the function has not been defined as the number of states.
 In fact, the condition (\ref{nonn}) in the P1 phase 
 has not been used to determine the phase boundary.
 The results in the P2 and SG phases may be understood 
 the number of states since the entropy is independent of 
 the imaginary part of the temperature.
 Therefore, although the definition of the entropy 
 in section~\ref{entropy} looks like a natural one,
 we need more discussion for the physical meaning of the entropy.

\end{itemize}

\section{GREM and the full RSB limit}
\label{gremb}

 In contrast to a naive expectation that the two-dimensional 
 distribution of zeros is specific to the SG phase,
 our result shows that zeros are not in the SG phase 
 but in a P phase.
 The area-distributed domain is not on the real axis and 
 cannot be directly related to the SG phase transition.

 The REM is considered the simplest SG model 
 since the SG phase is described by the 1RSB solution
 and the order parameter only takes either 0 or 1.
 We want to discuss more complicated situations
 keeping the advantage of the REM that
 the analytical calculation is possible.
 For this purpose, we employ 
 the GREM~\cite{Derrida4, DG1, DG2, OTT}.
 This model is a simple generalization of the REM and 
 allows us to analyze higher step RSB solutions easily.
 We consider the distribution of zeros in the GREM and 
 discuss possible distributions at the full RSB limit.

\subsection{Model}

 The GREM is defined by a hierarchical structure of energy levels.
 Each energy level is expressed by the sum of $K$ random numbers.
 To the $\nu$th level of hierarchy with $1\le\nu\le K$, 
 we assign random variables 
 $\epsilon_{\nu}(1), \epsilon_{\nu}(2), \cdots, \epsilon_{\nu}(M_\nu)$.
 The number of variables $M_\nu$ is given by 
\be
 M_\nu= (\alpha_1\cdots\alpha_\nu)^N, \label{mnu}
\ee
 where each $\alpha_\nu^N$ is an integer with 
 $1<\alpha_\nu^N< 2^N$ satisfying
\be
 M_K=(\alpha_1\cdots\alpha_K)^N=2^N. \label{alpha}
\ee
 For the $\nu$th level of hierarchy,
 we generate random numbers with Gaussian distribution
\be
 P_\nu(\epsilon_\nu) = \frac{1}{\sqrt{\pi NJ^2 a_\nu}}
 \exp\left(-\frac{\epsilon_\nu^2}{NJ^2 a_\nu}\right), 
\ee
 where $a_\nu>0$ satisfies 
\be
 \sum_{\nu=1}^K a_\nu = 1.
\ee
 From the random numbers generated as above,
 we construct $2^N$-random numbers as
\be
 E_i 
 = \sum_{\nu=1}^K\epsilon_{\nu}(\lfloor(i-1)M_\nu/2^N \rfloor+1),
 \label{Ei}
\ee
 where $i=1,2,\cdots, 2^N$ and $\lfloor x \rfloor$ is the floor function
 which indicates the largest integer not exceeding $x$.

 The GREM is defined as a system with energy levels (\ref{Ei}).
 By choosing parameters in an appropriate way, 
 we can find multiple phase transitions.
 We define 
\be
 T_\nu=\frac{J}{2}\sqrt{\frac{a_\nu}{\ln \alpha_\nu}}, 
 \label{Tnu}
\ee
 where $\nu=1,2,\cdots, K$.
 If $T_1>T_2>\cdots>T_K$,
 phase transitions occur at these temperatures.

 In the replica analysis of this model, 
 the SG order parameter $q_{ab}^{(\nu)}$
 is defined at each hierarchy $\nu$~\cite{OTT}.
 At $T>T_1$, 
 all the parameters are given by the P solution 
 $q_{ab}^{(\nu)}=\delta_{ab}$.
 At $T=T_\nu$, $q_{ab}^{(\nu)}$ turns into the 1RSB solution and 
 the degrees of freedom in the $\nu$th hierarchy fall into 
 the ground state.
 For example, when we consider the $K=2$ case,
 the system goes into the P-P, SG-P and SG-SG  
 phases as the temperature is decreased.
 It is understood from the complexity analysis that 
 the SG-P phase is identified as the 1RSB state and 
 the SG-SG phase as the two-step RSB one.

\subsection{Distribution of zeros}

\begin{center}
\begin{figure}[htb]
\begin{minipage}[h]{0.5\textwidth}
\begin{center}
\includegraphics[width=0.9\columnwidth]{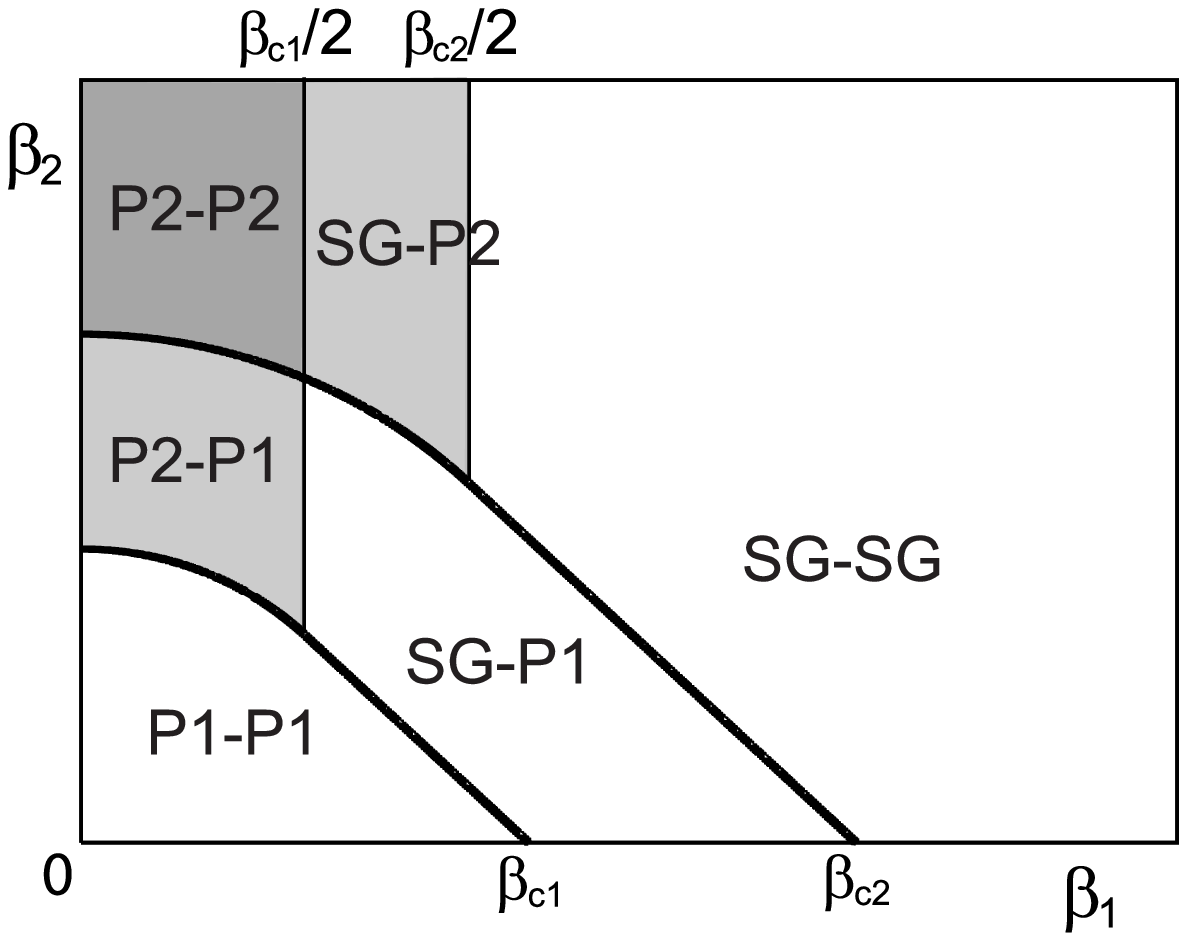}
\caption{Phase diagram and distribution of zeros 
 of the GREM with two hierarchies 
 in the complex-$\beta$ plane.
 The plot is only in the first quadrant.
 Zeros are distributed in the shaded regions and on the bold lines.
}
\label{grem}
\end{center}
\end{minipage}
\begin{minipage}[h]{0.5\textwidth}
\begin{center}
\includegraphics[width=0.9\columnwidth]{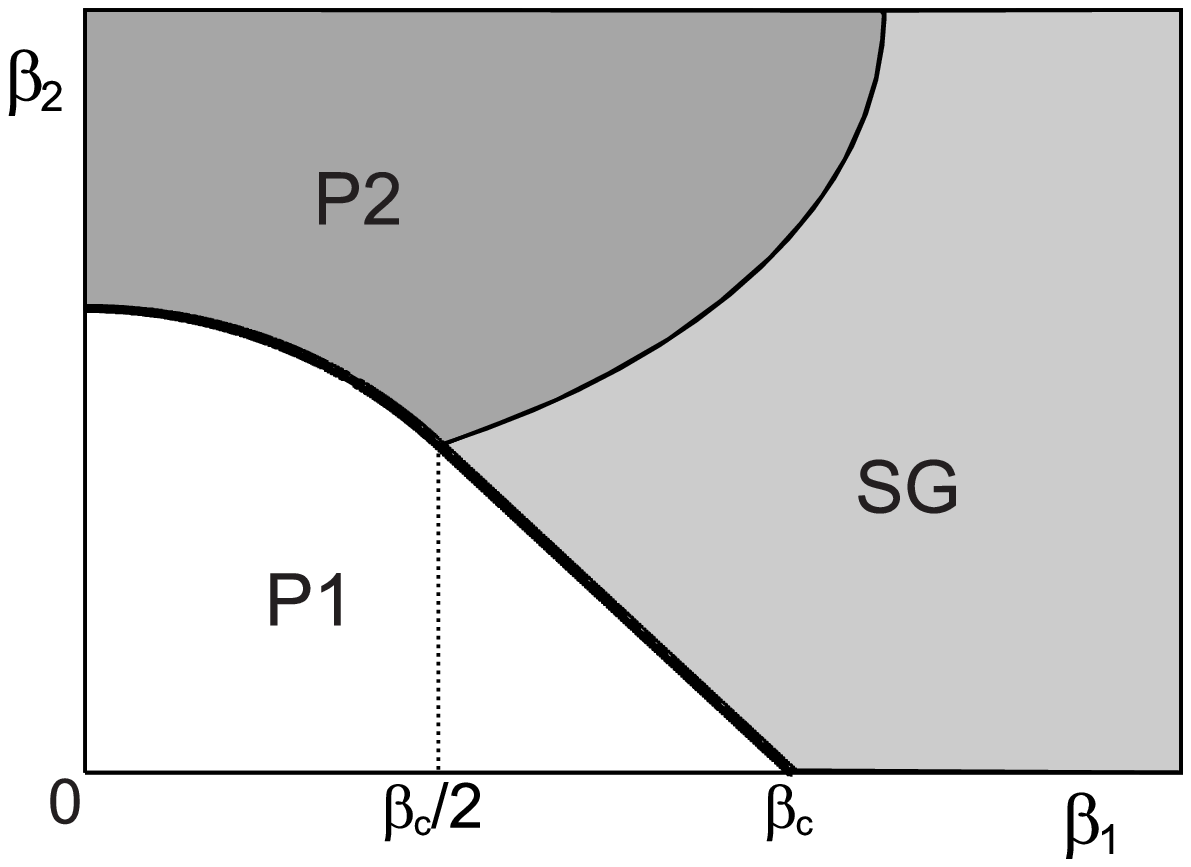}
\caption{A continuous limit $K\to\infty$ of the GREM.
Critical temperatures change continuously between
$T_1=1/\beta_{\rm c}$ and $T_K=0$.
}
\label{frsb}
\end{center}
\end{minipage}
\end{figure}
\end{center}

 It is a straightforward task to apply the calculation of zeros 
 in the previous section to the GREM.
 We consider two hierarchy ($K=2$) system and 
 choose parameters so that the relation $T_1>T_2$ is satisfied.
 We calculate $\phi=[\ln|Z|]/N$ in each phase.
 Following the calculation in \cite{OTT}, we find 
\be
 \phi = \left\{\ba{ll}
 \ln 2+\frac{1}{4}(\beta_1^2-\beta_2^2)J^2 & {\rm P1-P1} \\
 \frac{1}{2}\ln \alpha_1+\frac{a_1}{2}\beta_1^2J^2
 +\ln \alpha_2+\frac{a_2}{4}(\beta_1^2-\beta_2^2)J^2 & {\rm P2-P1} \\
 \frac{1}{2}\ln 2+\frac{1}{2}\beta_1^2J^2 & {\rm P2-P2} \\
 \beta_1J\sqrt{a_1\ln\alpha_1}
 +\ln \alpha_2+\frac{a_2}{4}(\beta_1^2-\beta_2^2)J^2 & {\rm SG-P1} \\
 \beta_1J\sqrt{a_1\ln\alpha_1}
 +\frac{1}{2}\ln \alpha_2+\frac{a_2}{2}\beta_1^2J^2 & {\rm SG-P2} \\
 \beta_1J\left(\sqrt{a_1\ln\alpha_1}+\sqrt{a_2\ln\alpha_2}\right) 
 & {\rm SG-SG} 
 \ea\right.,
\ee
 which is easily understood as a straightforward extension of 
 the result in the previous section to the two-hierarchy case.
 Then, comparing these solutions, we obtain 
 the phase diagram and distribution of zeros in figure~\ref{grem}.
 Zeros are distributed in the shaded regions and 
 on the bold lines of the figure.
 We have two phase transitions at $T=T_{1,2}$ and 
 correspondingly, two lines of zeros touch the real axis.
 Still, the shaded regions are far from 
 the real axis and are not related to the SG transition.

 An interesting result is obtained 
 when we consider the continuum limit of hierarchy.
 Then, the critical temperatures become continuous between 
 $T_1$ and $T_K$ and the SG phase can be understood 
 as the full step RSB state.
 We can find a similar, but different, behavior as 
 the Sherrington-Kirkpatrick model~\cite{SK} which also has 
 the full RSB SG phase. 
 In the GREM, the continuum limit is not unique 
 and depends on the choice of parameters~\cite{DG1}.
 We show a typical example with $T_K=0$ in figure~\ref{frsb}. 
 We see that the line distributions at phase boundaries 
 are accumulated to become area distributions.
 In the P2 phase of the figure, 
 the original area distributions and accumulated lines 
 overlap with each other and cannot be distinguished.
 The area distribution in the SG phase comes in contact with 
 the real axis, which is consistent with the picture 
 that phase transitions occur continuously 
 in the full RSB SG phase.
 Actually, a similar numerical result has been 
 obtained for a system on the Bethe lattice~\cite{MMNOS}.
 Our result in the GREM shows that 
 distributions of zeros in SG systems
 with a full RSB state are more complicated than those in the REM.

\section{Ferromagnetic interaction and first order transition}
\label{ferro}

\begin{center}
\begin{figure}[htb]
\begin{minipage}[h]{0.5\textwidth}
\begin{center}
\includegraphics[width=0.9\columnwidth]{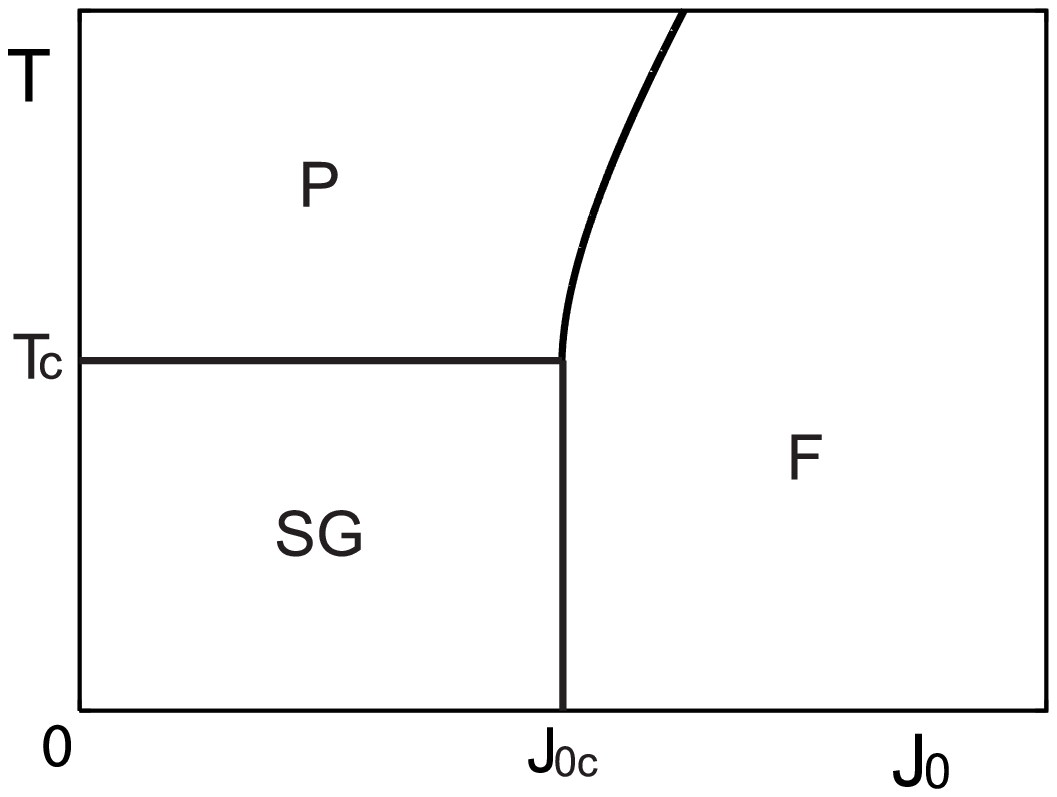}
\caption{Phase diagram of the REM with the F interaction $J_0$.
The P-F transition is of first order and
the others of second order.
}
\label{tj0}
\end{center}
\end{minipage}
\begin{minipage}[h]{0.5\textwidth}
\begin{center}
\includegraphics[width=0.9\columnwidth]{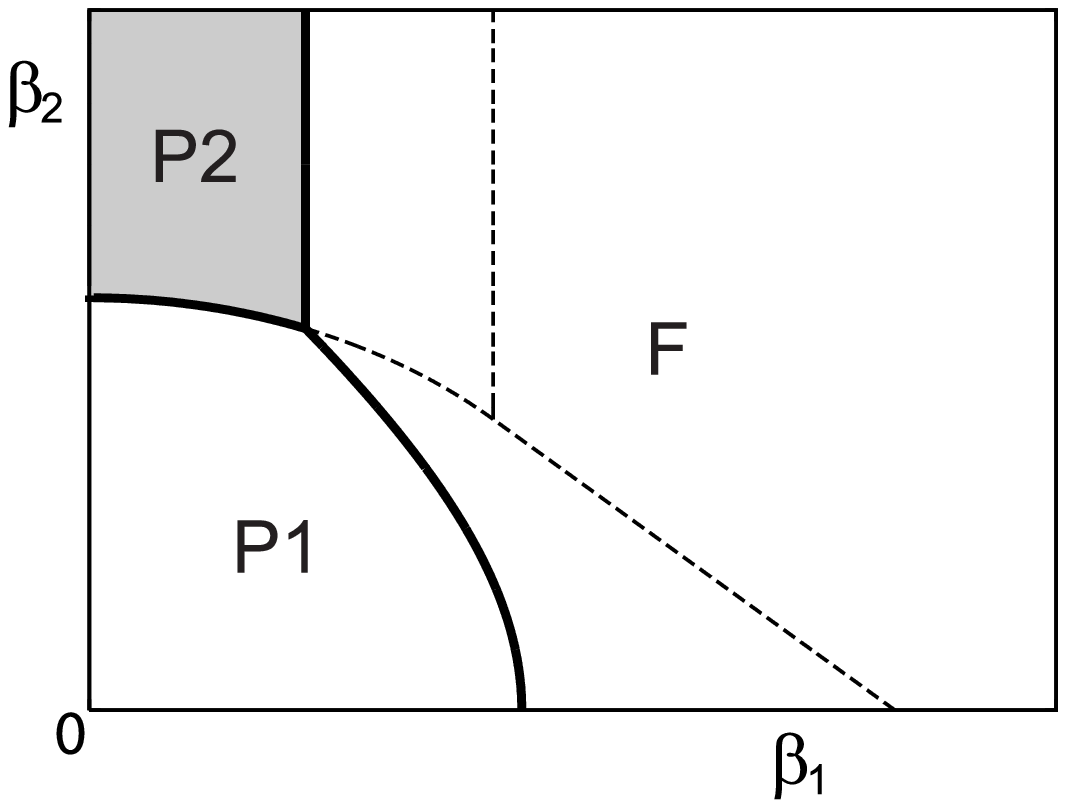}
\caption{Phase diagram and distribution of zeros of the REM 
at $J_0>J_{0{\rm c}}=J\sqrt{\ln 2}$.
The dashed lines are the phase boundaries at $J_0<J_{0\rm c}$.}
\label{j0}
\end{center}
\end{minipage}
\end{figure}
\end{center}

 In the previous sections, we have studied the REM.
 This model is described by a random distribution of energy levels
 and the spin degrees of freedom do not appear at all.
 Therefore, in order to extend our method to other spin models,
 it is necessary to treat the spin Hamiltonian.
 To achieve this, we treat the $p$-spin model in (\ref{pspin}). 
 At the same time, to motivate the use of the spin representation, 
 we treat the positive regular interaction $J_0$
 whose effect is incorporated to the probability distribution 
 of $J_{i_1\cdots i_p}$ as 
\be
 {\rm P}_{\rm J}(J_{i_1\cdots i_p})=
 \sqrt{\frac{N^{p-1}}{\pi p! J^2}}
 \exp\left\{
 -\frac{N^{p-1}}{p!J^2}
 \left(J_{i_1i_2\cdots i_p}-\frac{p!J_0}{N^{p-1}}\right)^2
 \right\}.
\ee
 At large $J_0$, the system is in 
 the ferromagnetic (F) phase which is characterized by 
 the magnetization $m=\sum_i\langle S_i\rangle/N$ where 
 $\langle\,\rangle$ denotes the thermal average.
 At the REM limit $p\to\infty$, the model can be solved exactly, 
 and the phase diagram in the $J_0-T$ plane is obtained 
 as in figure~\ref{tj0}.
 The phase boundary between 
 the SG and F phases 
 is given by $J_0=J_{0{\rm c}}=J\sqrt{\ln 2}$
 and that between the P and F phases by  
\be
 T = \frac{J^2}{2}\frac{1}{J_0-\sqrt{J_0^2-J^2\ln 2}}.
\ee
 The transition between the P and F phases is of first order and
 the entropy shows a discontinuous change.
 Therefore, by treating this model 
 we can study both the spin representation 
 and the first-order phase transition
 in our replica formulation.

 Following the standard method to calculate 
 $[Z^n]$~\cite{MPV, Nishimori, MM, GM}, 
 we can treat $[|Z|^{2n}]$ in a similar way.
 Since the replica space is doubled to 
 the complex conjugate space, 
 we have two kinds of spin variables $\{S_i^a\}$ and $\{S_i^{'a}\}$.
 Correspondingly, 
 the SG order parameters are defined as 
 $q_{ab}\sim \sum_i\langle S_i^aS_i^b\rangle/N$, 
 $q'_{ab}\sim \sum_i\langle S_i^{'a}S_i^{'b}\rangle/N$ and  
 $\tilde{q}_{ab}\sim \sum_i\langle S_i^aS_i^{'b}\rangle/N$. 
 Since we treat the F phase, we also have the magnetization 
 $m_{a}\sim \sum_i\langle S_i^a\rangle/N$  and 
 $m'_{a}\sim \sum_i\langle S_i^{'a}\rangle/N$.
 By using these variables, we can write 
\be
 [|Z|^{2n}] &=& \exp\left\{
 -N\frac{p-1}{2}\beta^2J^2\sum_{a>b}(q_{ab})^p
 -N\frac{p-1}{2}\beta^{*2}J^2\sum_{a>b}(q'_{ab})^p \right. \no\\
 & & 
 -N\frac{p-1}{2}|\beta|^2J^2\sum_{a>b}(\tilde{q}_{ab})^p
 +\frac{Nn\beta^2J^2}{4}+\frac{Nn\beta^{*2}J^2}{4} \no\\
 & & 
 -N(p-1)\beta J_0\sum_{a}(m_a)^p
 -N(p-1)\beta^* J_0\sum_{a}(m'_a)^p \no\\
 & & +N\ln\Tr \exp\left(
 \beta^2J^2\sum_{a>b}\hat{q}_{ab}S_aS_b
 +\beta^{*2}J^2\sum_{a>b}\hat{q}'_{ab}S'_aS'_b \right. \no\\
 & & \left.\left.
 +|\beta|^2J^2\sum_{a,b}\hat{\tilde{q}}_{ab}S_aS'_b
 +2\beta J_0\sum_{a}\hat{m}_aS_a
 +2\beta^* J_0\sum_{a}\hat{m}'_aS'_a
 \right)\right\},  \label{z2nj0}
\ee
 where variables with the hat symbol are related to those without hat as 
\be
 \hat{q}_{ab} = \frac{p}{2}(q_{ab})^{p-1}. \label{hat}
\ee
 The order parameters are obtained from the saddle-point equations.
 At $p\to\infty$, the equations are easily solved 
 since the order parameters take either 0 or 1.
 The P1, P2 and SG phases without the magnetization are calculated  
 in the same way as the previous case and
 (\ref{phi-rem}) is obtained.
 In the F phase given by $m_a=m'_a=1$, 
 all spins are aligned to the same direction and 
 we find a simple expression
\be
 \phi = \beta_1 J_0.
\ee

 Comparing $\phi$ in each phase, we can write the phase diagram
 and the distribution of zeros in the complex $\beta$ plane.
 When $J_0<J_{0{\rm c}}$, the F phase solution is irrelevant and
 we have the phase diagram in figure~\ref{rem}.
 The result at $J_0>J_{0{\rm c}}$ is shown in figure~\ref{j0}.
 The phase boundary between the P1 and F phases is given by 
\be
 \beta_2J = \sqrt{\beta_1^2J^2-4\beta_1J_0+4\ln 2} 
\ee
 and that between the P2 and F phases by 
\be
 \beta_1J^2 = J_0-\sqrt{J_0^2-J^2\ln 2}.
\ee
 The zeros are distributed in the P2 phase and 
 on all the phase boundaries.

 Comparing figures~\ref{rem} and \ref{j0}, 
 we can see the difference between the second- and first-order 
 phase transitions.
 The general consideration shows that 
 there is a relation between the order of the phase transition
 and the distributions of zeros on the real axis~\cite{Abe}.
 In the second-order transition, 
 the line distribution meets with the real axis with the angle of
 $\pi/4$, and the density of zeros goes to zero on the real axis.
 In contrast, for the first-order transition,
 the angle between the line and the real axis is $\pi/2$
 and the density is finite on the real axis.
 Our results support these expectations.
 On the real axis, the density of zeros is given by 
\be
 \rho_\beta(\beta_1,\beta_2=0) &=& 
 \frac{J^2}{2\pi}\sqrt{J_0^2-J^2\ln 2}
 \ \theta(J_0-J\sqrt{\ln 2}) 
 \no\\
 & & \times\delta\left(\beta_1J^2-2(J_0-\sqrt{J_0^2-J^2\ln 2})\right).
 \label{deltae}
\ee
 It is also known from the general argument that  
 the density at the transition point is directly related to 
 the discontinuous energy change $\Delta \epsilon$.
 In the present case, it is obtained as 
 $\Delta\epsilon=\epsilon_{\rm P}-\epsilon_{\rm F}
 =\sqrt{J_0^2-J^2\ln 2}$.
 We see that the same factor appears in (\ref{deltae}).

\section{Lee-Yang zeros of the REM}
\label{mag}

 Up to here we have discussed the Fisher zeros in the REM.
 In this section, we treat the Lee-Yang zeros in the same model.
 While the analysis of systems with complex temperature 
 clarifies the system's thermodynamic properties 
 such as the entropy and specific heat, 
 that with complex magnetic field is useful to understand 
 the magnetic properties such as the magnetization and susceptibility.

 The analysis of the Fisher zeros using the replica method 
 goes along the same line as that of the Lee-Yang zeros 
 in the previous sections.
 To apply the magnetic field, we consider two possible patterns: 
 longitudinal and transverse magnetic fields. 
 The density of zeros in the complex longitudinal field 
 has been studied in \cite{MP2,MP3}.
 The method used there is very similar to 
 the original analysis of the Fisher zeros in the REM~\cite{Derrida3}.
 We re-derive the result by using the replica method.
 Also, we treat the system with a transverse field, 
 which allows us to study the quantum fluctuation effect.

\subsection{Longitudinal magnetic field}

 We use the energy representation of the REM 
 in (\ref{Z}) and (\ref{gauss}).
 Then, the magnetic field $h$ is 
 incorporated to the partition function as 
\be
 Z = \sum_{M}e^{\beta hM}\sum_{i=1}^{N(M)}e^{-\beta E_i(M)}, 
\ee
 where $M$ represents the ``magnetization'' and takes 
\be
 M_k = -(N-2k), 
\ee
 with $k=0,1,\cdots,N$.
 For a given $M_k$, the number of the sum $N(M_k)$ 
 is equal to $N!/k!(N-k)!$.
 The partition function is written as 
\be
 Z = e^{-N\beta h}\sum_{k=0}^{N} y^k
 \sum_{i=1}^{N(M_k)} e^{-\beta E_{ki}},
\ee
 where $y=e^{2\beta h}$ and 
 $E_{ki}$ is a random variable taken from (\ref{gauss}).
 Thus, the partition function is a polynomial of the $N$th degree 
 in $y$ and the density of zeros is obtained from (\ref{rho}).

\begin{center}
\begin{figure}[htb]
\begin{minipage}[h]{0.5\textwidth}
\begin{center}
\includegraphics[width=0.9\columnwidth]{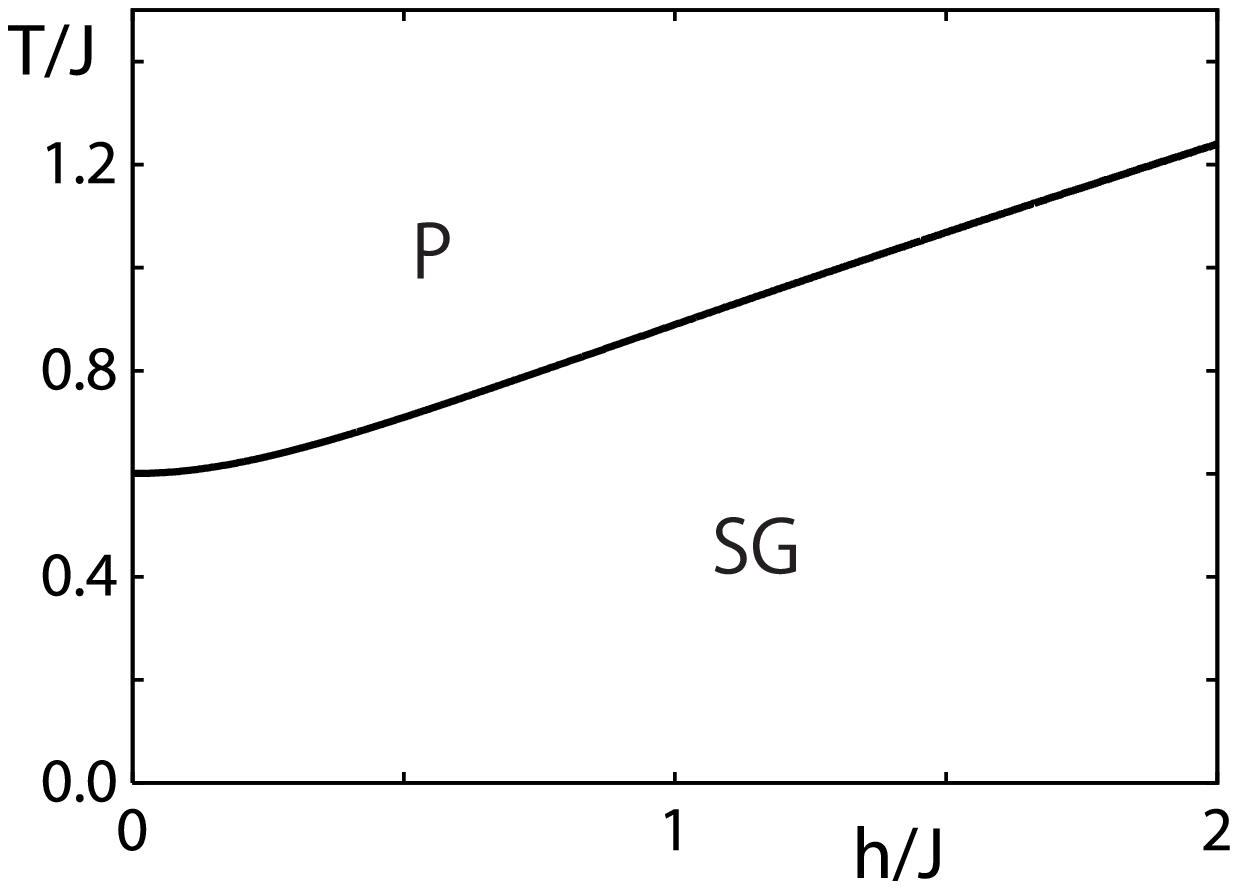}
\caption{Phase diagram of the REM in the $h-T$ plane.
The P-SG transition is of second order.
}
\label{th}
\end{center}
\end{minipage}
\begin{minipage}[h]{0.5\textwidth}
\begin{center}
\includegraphics[width=0.9\columnwidth]{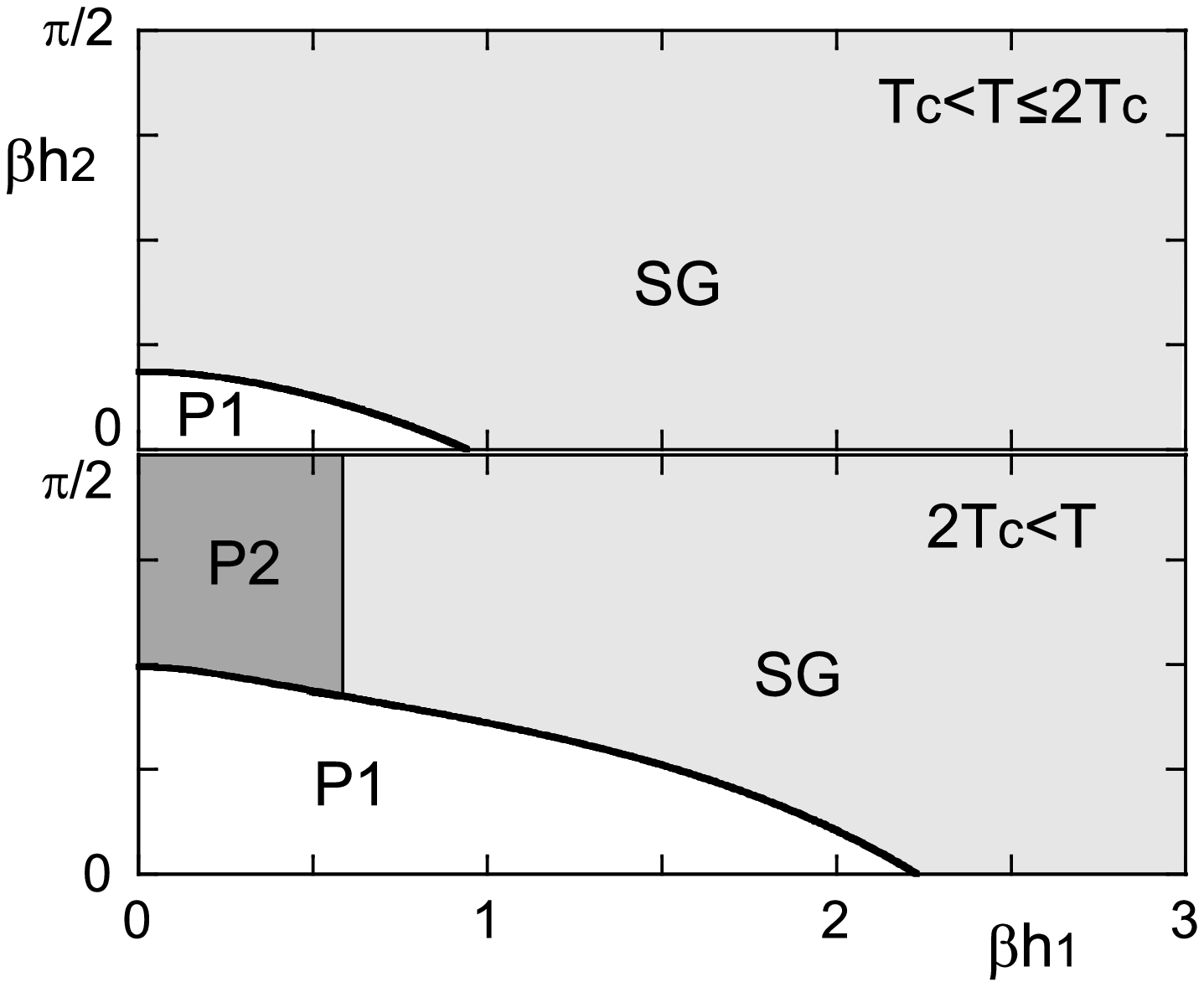}
\caption{Phase diagram and distribution of zeros of the REM
in the complex-$h$ plane. 
The upper figure is for $T_{\rm c}<T\le 2T_{\rm c}$
and lower for $2T_{\rm c}<T$.
For $T\le T_{\rm c}$, all domain belongs to the SG phase.
}
\label{h}
\end{center}
\end{minipage}
\end{figure}
\end{center}

 The REM with real $h$ has been treated in \cite{Derrida1, Derrida2} 
 and the phase diagram is depicted in figure~\ref{th}.
 The P and SG phases are separated by the critical temperature 
 $T_{\rm c}(h)=1/\beta_{\rm c}(h)$ defined by 
\be
 \frac{\beta_{\rm c}^2(h)J^2}{4}
 =\ln(1+e^{2\beta_{\rm c}(h)h})
 -\frac{2\beta_{\rm c}(h)h }{1+e^{-2\beta_{\rm c}(h) h}}.
\ee
 As in the previous cases, 
 this transition is also caused by the entropy crisis 
 which means that the entropy goes to zero in the SG phase.

 When $h$ is a complex number, 
 the P1, P2 and SG phases appear in the phase diagram.
 $|Z|^{2n}$ is written as
\be
 |Z|^{2n} &=& e^{-2Nn\beta h_1}
 \sum_{\{k_a,i_a\}}\sum_{\{k'_a,i_a'\}}
 (y)^{\sum_{k=0}^N kn_k(\{k_a\})}(y^*)^{\sum_{k=0}^N kn'_k(\{k'_a\})}
 \no\\
 & & \times
 \exp\left\{
 -\beta \sum_{k=0}^N\sum_{i=1}^{N(M_k)}
 \left(n_{ki}(\{k_a,i_a\})+n'_{ki}(\{k'_a,i'_a\})\right)E_{ki}
 \right\},
\ee
 where 
\be
 & & n_{ki}(\{k_a,i_a\}) = \sum_{a=1}^n
 \delta_{k_ak}\delta_{i_ai}, \\
 & & n_{k}(\{k_a\}) =  \sum_{a=1}^n \delta_{k_ak}.
\ee
 Each $k_a$ takes $0,1,\cdots,N$ and 
 $i_a$ takes $1,2,\cdots,N(M_{k_a})$.
 Taking the average, we obtain 
\be
 [|Z]^{2n} &=& e^{-2Nn\beta h_1}
 \sum_{\{k_a,i_a\}}\sum_{\{k'_a,i_a'\}}
 (y)^{\sum_{k=0}^N kn_k(\{k_a\})}(y^*)^{\sum_{k=0}^N kn'_k(\{k'_a\})}
 \no\\
 & & \times
 \exp\left\{
 \frac{N\beta^2J^2}{4}\sum_{a,b=1}^n
 \left(q_{ab}+q'_{ab}+2\tilde{q}_{ab}\right)
 \right\},
\ee
 where the order parameters are defined as 
\be
 & & q_{ab} = \delta_{k_ak_b}\delta_{i_ai_b}, \\
 & & q'_{ab} = \delta_{k_ak_b}\delta_{i'_ai'_b}, \\
 & & \tilde{q}_{ab} = \delta_{k_ak_b}\delta_{i_ai'_b}.
\ee
 As we mentioned above, we have three phases.
\begin{itemize}
\item{P1.}

 The simplest solution is given by 
 $q_{ab}=q_{ab}'=\delta_{ab}$ and $\tilde{q}_{ab}=0$. 
 We have for $\phi=[\ln|Z|]/N$ 
\be
 \phi_{\rm P1} = -\beta h_1+\frac{\beta^2J^2}{4}
 +\frac{1}{2}\ln(1+e^{2\beta h}) +\frac{1}{2}\ln(1+e^{2\beta h^*}).
\ee
 This is a simple extension 
 of the usual P phase result for real $h$ 
 to the case of complex $h$.

\item{P2.}

 The second simplest case is specific to the complex parameters:
 $q_{ab}=q_{ab}'=\tilde{q}_{ab}=\delta_{ab}$.
 Then, we  have the P2 phase result
\be
 \phi_{\rm P2} = -\beta h_1+\frac{\beta^2J^2}{2}
 +\frac{1}{2}\ln(1+e^{4\beta h_1}).
\ee
 This is valid at 
 $\beta \le \tilde{\beta}_{\rm c}(h_1)=\beta_{\rm c}(h_1)/2$ 
 where the entropy is nonnegative.

\item{SG.}

 When the temperature is  lower than $1/\tilde{\beta}_{\rm c}(h_1)$,
 the system freezes to the SG phase.
 The order parameters are given by 
 (\ref{1rsb-q}) and (\ref{1rsb-qt}).
 Then, 
\be
 \phi_{\rm SG} = -\beta h_1
 +\frac{\beta\tilde{\beta}_{\rm c}(h_1)J^2}{2}
 +\frac{\beta}{2\tilde{\beta}_{\rm c}(h_1)}
 \ln(1+e^{4\tilde{\beta}_{\rm c}(h_1)h_1}).
\ee
\end{itemize}
 
 Comparing these results, we can draw the phase diagram 
 in the complex-$h$ plane as figure~\ref{h}
 \footnote{The phase diagram should in principle be drawn 
 in the complex-$y$ plane, 
 though the translation between $y$ and $h$ is immediate.
 Then, the $h$-plane is restricted to
 $|2\beta h_2|\le \pi$.}.
 Zeros can appear in the P2 and SG phases, 
 and the P1-P2 and P1-SG boundaries.
 The result is completely the same as 
 that in the previous study obtained by a different method~\cite{MP2}.

 It is interesting to see that the density of zeros in the SG phase 
 is nonzero as 
\be
 \rho_h(h_1,h_2)=\beta\tilde{\beta}_{\rm c}(h_1)
 \frac{1-\tanh^2(2\tilde{\beta}_{\rm c}(h_1)h_1)}
 {1+2(h_1^2/J^2)(1-\tanh^2(2\tilde{\beta}_{\rm c}(h_1)h_1))},
 \label{doz}
\ee
 in contrast to the case of the Fisher zeros where 
 the zeros do not appear in the SG phase.
 In the present case, although the entropy is zero in the SG phase, 
 the critical temperature depends on the magnetic field
 and the value of the free energy depends sensitively on the field.
 As a result, the density of zeros gives a finite value.
 But we note that the function (\ref{doz}) decays exponentially 
 in $h_1$ and takes considerably small values,
 which mean that zeros very rarely exist in the SG phase.

 Another interesting point is that the P2 phase 
 appears at $T\ge 2T_{\rm c}=1/\tilde{\beta}_{\rm c}(0)$.
 This property is the same as that of the Fisher zeros 
 (see figures~\ref{rem}, \ref{grem} and \ref{frsb}).
 At present, the reason why $2T_{\rm c}$ 
 plays a special role is not clear. 
 It may be interesting to see 
 whether this property is specific to the REM or not.
 
\subsection{Transverse magnetic field}

\begin{center}
\begin{figure}[htb]
\begin{minipage}[h]{0.5\textwidth}
\begin{center}
\includegraphics[width=0.9\columnwidth]{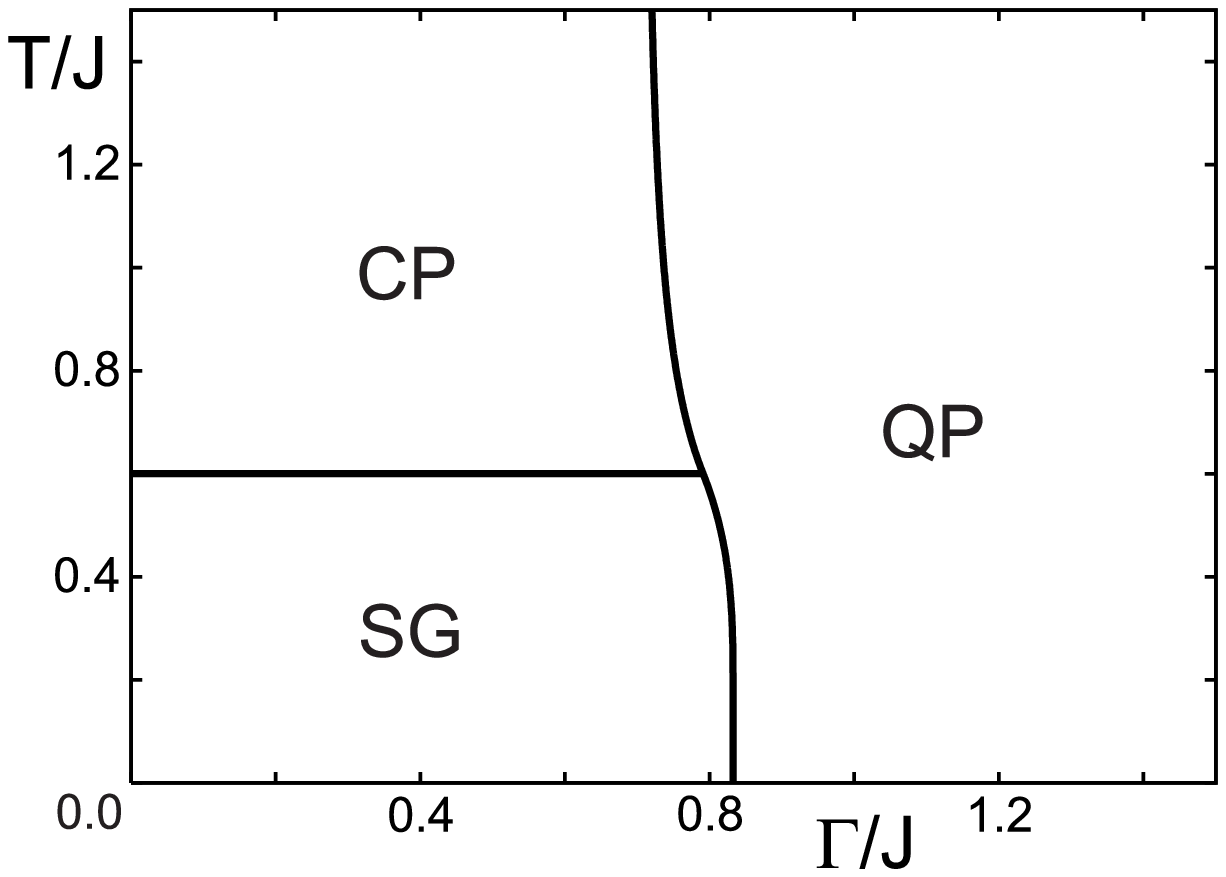}
\caption{Phase diagram of the REM in a transverse magnetic field $\Gamma$.
The CP-QP and SG-QP transitions are of first order.
}
\label{tg}
\end{center}
\end{minipage}
\begin{minipage}[h]{0.5\textwidth}
\begin{center}
\includegraphics[width=0.9\columnwidth]{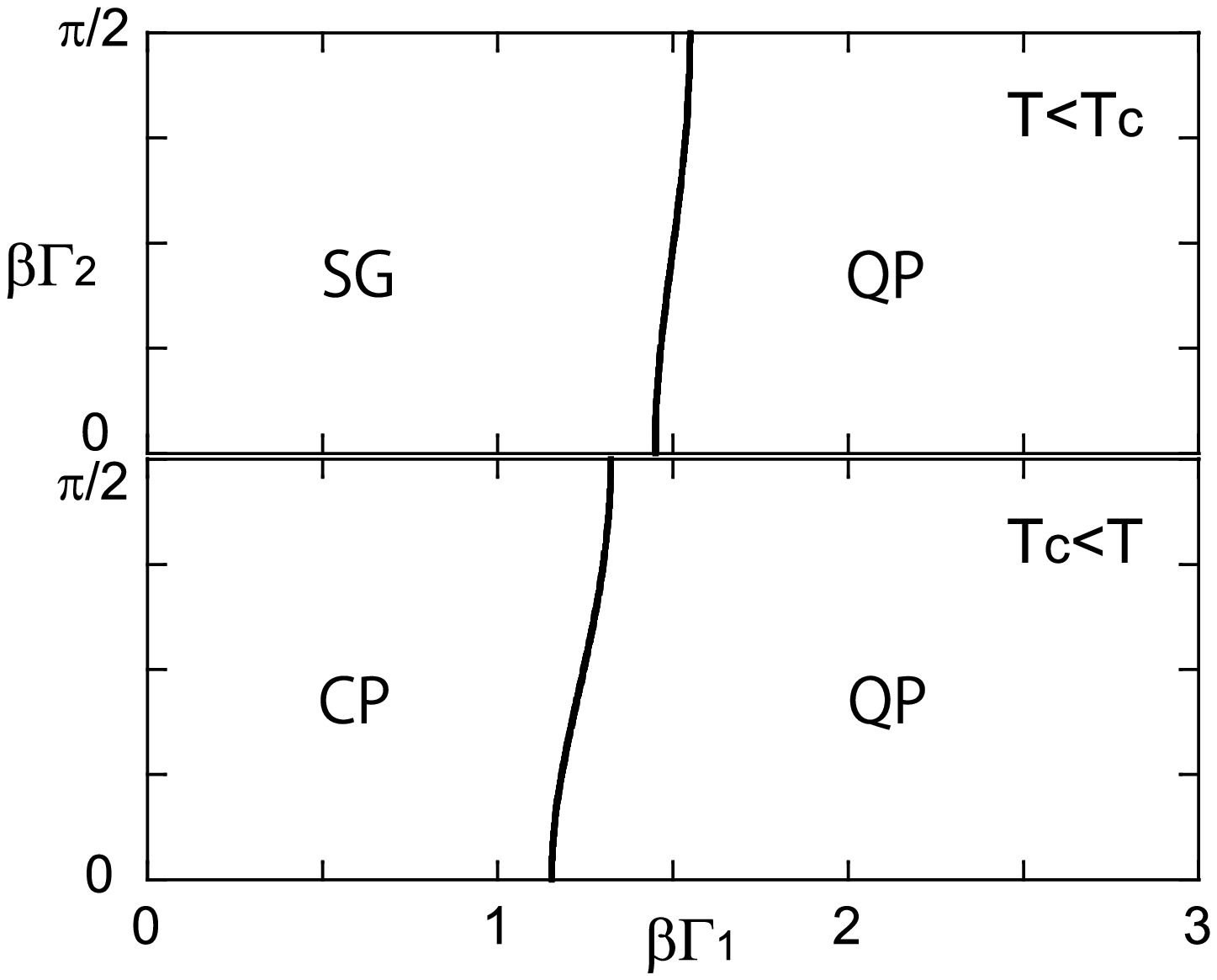}
\caption{Phase diagram and distribution of zeros of the REM
in the complex-$\Gamma$ plane. 
The upper figure is for $T<T_{\rm c}=J/2\sqrt{\ln 2}$
and lower for $T_{\rm c}<T$.
}
\label{g}
\end{center}
\end{minipage}
\end{figure}
\end{center}

 Then, we consider the REM in a transverse field~\cite{Goldschmidt}.
 The Hamiltonian is given by 
\be
 H= -\sum_{i_1<\cdots<i_p}J_{i_1\cdots i_p}
 \sigma_{i_1}^z\cdots \sigma_{i_p}^z
 -\Gamma\sum_{i=1}^N\sigma_i^x,
\ee
 where $\sigma_z$ and $\sigma_x$ are the Pauli matrices.
 Since two terms in the right-hand side of the equation 
 do not commute with each other, 
 the system must be treated quantum mechanically.
 At the REM limit $p\to \infty$, this model can be solved exactly and 
 the phase diagram is given by figure~\ref{tg}.
 We have two kinds of P phases:
 the classical P (CP) and quantum P (QP) phases.
 The CP phase is the same as the P phase in the classical REM
 and spins are directed along the $z$-direction randomly.
 On the other hand, in the QP phase,
 spins are aligned with 
 the transverse-field direction.

 The average replicated partition function is written 
 in terms of order parameters
 $\chi_a(\tau,\tau')\sim 
 \sum_i\langle S_{zi}^a(\tau)S_{zi}^a(\tau')\rangle/N$
 and 
 $q_{ab}(\tau,\tau')\sim 
 \sum_i\langle S_{zi}^a(\tau)S_{zi}^b(\tau')\rangle/N$ 
 where $a\ne b$.
 Since we are treating the quantum system, 
 the trace is defined by the imaginary time path integral with some
 measure~\cite{Takahashi}, and 
 the spin variable $\mbox{\boldmath $S$}_i^a$ depends on the time $\tau$.
 The order parameter $\chi_a$ with $q_{ab}\le \chi_a\le 1$ 
 characterizes quantum effects 
 since $\chi_a=1$ in the classical system.

 When $\Gamma$ is real, 
 the model is solved by the 1RSB and static ansatz.
 We expect that the same ansatz can be applied 
 to the case of complex $\Gamma$.
 Proceeding in a similar way as the previous cases, 
 we have 
\be
 [|Z|^{2n}] &=& \exp\left\{
 -\frac{Nn\beta^2J^2}{4}(p-1)
 \Bigl(\chi^p-(1-m)q^p\Bigr) \right.\no\\
 & & \qquad
 -\frac{Nn\beta^{*2}J^2}{4}(p-1)
 \Bigl(\chi'^p-(1-m)q'^p\Bigr) \no\\
 & & \left. \qquad
 -\frac{Nn|\beta|^2J^2}{2}(p-1)\Bigl(
 \tilde{\chi}^p-(1-m)\tilde{q}^p\Bigr)
 +N\ln\Tr e^L \right\}, 
 \\
 \Tr e^L &=& \Tr\exp\left\{
 \frac{\beta^2J^2}{2}\left(\hat{\chi}-\hat{q}\right)
 \sum_{a=1}^n(\hat{S}_z^a)^2
 +\frac{\beta^2J^2}{2}\hat{q}\sum_{\rm B}^{n/m}\left(
 \sum_{a\in B}^m \hat{S}_z^a\right)^2
 \right.\no\\
 & & 
 +\frac{\beta^2J^2}{2}\left(\hat{\chi}'-\hat{q}'\right)
 \sum_{a=1}^n(\hat{S}_z^{'a})^2
 +\frac{\beta^2J^2}{2}\hat{q}'\sum_{\rm B}^{n/m'}\left(
 \sum_{a\in B}^{m'} \hat{S}_z^{'a}\right)^2
 \no\\
 & & 
 +\beta^2J^2(\hat{\tilde{\chi}}-\hat{\tilde{q}})
 \sum_{a=1}^n\hat{S}_z^{a}\hat{S}_z^{'a}
 +\frac{\beta^2J^2}{2}\hat{\tilde{q}}\sum_{\rm B}^{n/m}
 \left(\sum_{a\in B}^m \hat{S}_z^a\right)
 \left(\sum_{a\in B}^m \hat{S}_z^{'a}\right)
 \no\\
 & & \left.
 +\beta\Gamma\sum_{a=1}^n\hat{S}_x^a
 +\beta\Gamma^*\sum_{a=1}^n\hat{S}_x^{'a}
 \right\},
\ee
 where 
 $\hat{S}^a_z=\int_0^\beta  d\tau\, S^a_z(\tau)/\beta$ 
 and the order parameters with the hat symbol 
 are defined in the same way as in (\ref{hat}).
 Possible saddle-point solutions are given as follows.
\begin{itemize}
\item{CP.}

 $\chi=\chi'=1$ and the other parameters are set to zero.
 In this case the quantum effect is irrelevant, and we have 
 the CP result
\be
 \phi_{\rm CP} = \frac{\beta^2J^2}{4}+\ln 2.
\ee 
 From the nonnegative condition of the entropy, 
 this solution is shown to be effective at 
 $T>T_{\rm c} = J/2\sqrt{\ln 2}$.

\item{QP.}

 When the transverse field is strong enough so that
 the effect of interaction is negligible, 
 there is not any order in the $z$-direction. 
 Then, all the order parameters are neglected, and 
 we obtain 
\be
 \phi_{\rm QP} = 
 \frac{1}{2}\ln(e^{\beta\Gamma}+e^{-\beta\Gamma})
 +\frac{1}{2}\ln(e^{\beta\Gamma^*}+e^{-\beta\Gamma^*}).
\ee

\item{SG.}

 At low temperatures the system freezes to the ground state.
 In this case, $\chi=q=1$, and we have the standard 1RSB result
\be
 \phi_{\rm SG} = \beta J\sqrt{\ln 2}.
\ee
\end{itemize}

 We see that 
 any phase which is specific to the complex parameters 
 does not appear.
 For example, we can examine the CP2 phase defined by 
 $\chi=\chi'=\tilde{\chi}=1$ and $q=q'=\tilde{q}=0$.
 Then, we have 
\be
 \phi_{\rm CP2} = \frac{\beta^2J^2}{2}+\frac{1}{2}\ln 2.
\ee
 This solution is valid at $T>2T_{\rm c} = J/\sqrt{\ln 2}$.
 In this region, 
 $\phi_{\rm CP2}$ is smaller than $\phi_{\rm CP}$.
 Therefore, we can conclude that the CP2 phase 
 is irrelevant.
 In the same way, 
 we can also show that the ``QP2'' phase does not appear.

 The phase diagram is shown in figure~\ref{g}.
 We do not have any two-dimensional distribution of zeros
 and zeros are distributed on the phase boundary.
 The phase boundaries between CP/SG and QP phases are
 determined by 
\be
 \cosh(2\beta\Gamma_1)+\cos(2\beta\Gamma_2) 
 = \frac{1}{2} e^{2\phi_{\rm CP/SG}}.
\ee
 Their transitions are of first order, and we see that 
 the general argument, 
 the lines of zeros are vertical to the real axis, 
 also holds in this quantum case.

\section{Conclusions}
\label{conc}

 We have discussed the partition-function zeros 
 in random energy models by using a new replica method.
 We re-derived previously obtained results and 
 obtained several new results
 on the distribution of zeros of random systems.
 Our method is very general and systematic and  
 can in principle be applied to any spin model.

 We find that the two-dimensional distribution of zeros is
 characterized by the order parameter $\tilde{q}_{ab}$ 
 which connects configurations between 
 the original and complex conjugated spaces. 
 In the REM with complex temperature, 
 the area-distributed phase is paramagnetic
 rather than a SG phase.
 It does not appear at real $\beta$, 
 but the physical quantities of the real system are affected 
 by this phase any way 
 as we have demonstrated for the specific heat.

 We also find differences between first- and second-order 
 transitions.
 We showed that the general argument discussed in \cite{Abe} holds 
 even for random systems.
 Since our result is not a rigorous proof of the statement, 
 it would be interesting to find a counterexample in other models.

 In contrast to previous numerical analyses 
 of finite-dimensional SG systems, 
 we find in the REM that 
 the area distribution of zeros does not approach the real axis.
 This is because the REM is an oversimplified SG model.
 In the REM, the SG phase is described by the 1RSB ansatz and 
 the entropy goes to zero in the SG phase.
 These properties do not hold in general.
 In fact, the analysis of the GREM at the continuum limit 
 shows a more complicated behavior.
 Therefore, the next thing to do will be to apply our method 
 to other mean-field models 
 such as the Sherrington-Kirkpatrick model~\cite{SK}
 and the spherical model~\cite{KTJ, CS}. 
 The formulation is the same as in the present case, and 
 we can utilize the expression 
 (\ref{z2nj0}) for general $p$-spin systems.
 The difference arises 
 when we solve the saddle-point equation.
 The SG order parameters 
 take values between 0 and 1, and possibly complex values.
 Such a situation does not exist for the REM, and 
 we need a careful analysis,
 which will be done in a future work.

 Another point to be discussed is to see how 
 the thermodynamic limit is obtained 
 by taking the size of the system to infinity.
 We want to know how the zeros approach the real axis of the complex plane.
 Our analysis is based on the saddle-point method and 
 is valid only in the thermodynamic limit.
 Therefore, we need to take a different way.
 Recently, the spherical model has been solved 
 for arbitrary system size~\cite{AR}.
 If the method developed there is also useful 
 for systems with complex parameters, 
 we can study finite size effects for the distribution of zeros.
 It will be an interesting task to solve that problem
 in future works.

\section*{Acknowledgments}

 The author is grateful to Y Matsuda, H Nishimori, T Obuchi and K Takeda 
 for useful discussions and comments. 


\end{document}